\documentclass[aps,prl,reprint,superscriptaddress]{revtex4-1} 
\usepackage{hyperref} 
\usepackage{graphicx}% Include figure files
\usepackage{bm}% bold math
\begin{document}

%\preprint{}

\title{Single-neuron criticality optimizes analog dendritic computation}
%\title{The critical single-neuron hypothesis for dendritic computation}
%\title{The critical neuron hypothesis for dendritic computation}
%\title{Single neuron as a critical excitable medium}
%\title{The critical neuron conjecture: a single neuron can be a critical excitable medium}

%\title{[Statistical Physics approach to dendritic computation]{A
%  Statistical Physics approach to dendritic computation: the
%  excitable-wave mean-field approximation}}



\author{Leonardo L. Gollo}%
\email{leonardo.l.gollo@gmail.com}
%\thanks{corresponding author}
\affiliation{
IFISC, Instituto de F\'{\i}sica Interdisciplinar y Sistemas 
Complejos (CSIC - UIB), Campus Universitat de les Illes Balears, E-07122 Palma de Mallorca, 
Spain
}%
\affiliation{Systems Neuroscience Group, Queensland Institute of Medical Research, Brisbane, QLD 4006, 
Australia}
%\address{
%Laborat\'orio de F{\'\i}sica Te\'orica e Computacional, Departamento
%de F{\'\i}sica, Universidade Federal de Pernambuco, 50670-901 Recife,
%PE, Brazil}%

\author{Osame Kinouchi}%
 \email{osame@ffclrp.usp.br}
\affiliation{
Faculdade de Filosofia, Ci\^encias e Letras de Ribeir\~ao Preto,
Universidade de S\~ao Paulo, Avenida dos Bandeirantes 3900, 14040-901,
Ribeir\~ao Preto, SP, Brazil}%
\affiliation{
Center for Natural and Artificial Information Processing Systems - USP}

\author{Mauro Copelli}%
\email{mcopelli@df.ufpe.br} 
\affiliation{Departamento de F{\'\i}sica,
  Universidade Federal de Pernambuco, 50670-901 Recife, PE, Brazil}%
\affiliation{
Center for Natural and Artificial Information Processing Systems - USP}

%\author{}
%\thanks{}
%\affiliation{}


\date{\today}

\begin{abstract}
Neurons are thought of as the building blocks of excitable brain tissue. However, at the single neuron level, the neuronal membrane, the dendritic arbor and the axonal projections can also be considered an extended active medium. Active dendritic branchlets enable the propagation of dendritic spikes, whose computational functions, despite several proposals, remain an open question. Here we propose a concrete function to the active channels in large dendritic trees. By using a probabilistic cellular automaton approach, we model the input-output response of large active dendritic arbors subjected to complex spatio-temporal inputs and exhibiting non-stereotyped dendritic spikes. We find that, if dendritic spikes have a non-deterministic duration, the dendritic arbor can undergo a continuous phase transition from a quiescent to an active state, thereby exhibiting spontaneous and self-sustained localized activity as suggested by experiments. Analogously to the critical brain hypothesis, which states that neuronal networks self-organize near a phase transition to take advantage of specific properties of the critical state, here we propose that neurons with large dendritic arbors optimize their capacity to distinguish incoming stimuli at the critical state. We suggest that ``computation at the edge of a phase transition'' is more compatible with the view that dendritic arbors perform an analog rather than a digital dendritic computation. 
\end{abstract}

\maketitle

Critical systems are organized in a fractal-like pattern spanning
across numerous temporal and spatial scales. 
The brain has recently been included amongst abundant physical and
biological systems exhibiting traces of
criticality~\citep{Chialvo10,Sornette00}.  
In the past decade, several phenomena suggestive of critical states
have been observed in different systems: multielectrode data from
cortical slices {\it in vitro}~\citep{Beggs03, Beggs04,Friedman12},
anesthetized~\citep{Gireesh08}, awake~\citep{Petermann09} and behaving
animals~\citep{Ribeiro10}; human
electrocorticography~\citep{Miller09},
electroencephalography~\citep{Linkenkaer01},
magnetoencephalography~\citep{Kitzbichler09,Linkenkaer01}, as well as
functional magnetic resonance imaging~\citep{Eguiluz05,Kitzbichler09}
recordings.  Together, these experiments suggest evidence for the
intrinsic pervasiveness of criticality into a wide range of
spatio-temporal brain scales.

The critical brain hypothesis offers an appealing solution to a
long-lasting conundrum of neuroscience: how can localized information
(sometimes from very specific regions) propagate in the brain without
a spatial/temporal exponential decay or
saturation~\citep{Chialvo10,Sporns10}?  Advantages of neuronal systems
poised around a critical state include optimization of the dynamic
range of neuronal networks~\citep{Kinouchi06}, as confirmed
experimentally ~\citep{Shew09}, as well as transmission and storage of
information~\citep{Beggs03, Beggs04,Haldeman05,Shew11}. However, a
fundamental assumption of the critical brain hypothesis has so far not
been examined: is the single neuron the minimal dynamic unit in
neuronal networks, like a spin site in the Ising model for
ferromagnets (the prototypical system for phase transitions)? Or can
critical phenomena associated with phase transitions already occur at
the single neuron level?  Adopting a statistical physics approach, we
explore the behavior of dendritic branchlets at the sub-cellular level
as forming a network of functional dynamical units.

\begin{figure}%[!ht]
\begin{center}
\includegraphics[angle=0,width=1.0\columnwidth]{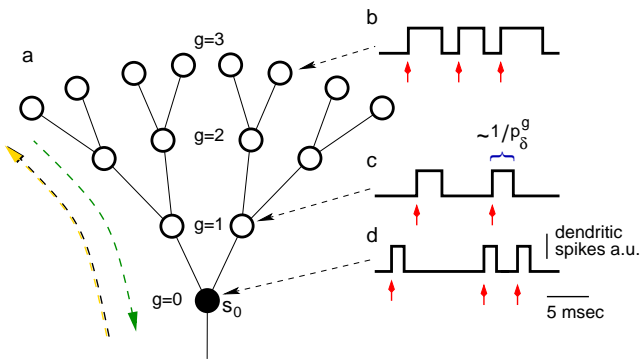}
\caption{\label{figDendrite} {\bf Model of an active dendritic tree
    with non-stereotyped dendritic spikes}. %- variable duration}.
  {\bf (a)}, Excitable elements (circles) connected (bars) in a Cayley
  tree topology with $G=3$ layers and coordination number $z=3$ (one
  mother and $k=2$ daughter branches). 
  Large green (yellow) arrow illustrates forward (backward) propagation.
  {\bf (b)-(d)}, Biologically
  motivated dendritic spike with duration depending on the
  distance from the soma as a net effect of variable density of ionic
  channels. Red arrows indicate that the site is being stimulated (by
  an external input and/or by mother or daughter branches), and 
  $\frac{1}{p_{\delta}^{g}}$ represents the average period of dendritic 
  spikes at layer $g$.}
\end{center}
\end{figure}

% The quest for the basic neuronal dynamic unit is conceptually
% important, but has been yet unexplored, probably because it requires
% an unusual perspective.  
Here we treat neurons with their extensive dendritic trees as
excitable media (see Fig.~\ref{figDendrite}a).  This task is far from
trivial because of the complexity~\citep{Wen09,Zomorrodi10}, the
diversity of neurons~\citep{Snider10}, and the absence of information
about key elements governing the dynamics, such as how the zoo of
ionic channels is distributed along the
dendrites~\citep{Koch,Reyes01,Johnston08,Coop10}.  In fact, there is
an entire research area concerned with biophysically detailed neuronal
modeling~\citep{Rall64,Neuron,Stuart99,Coop10} and dendritic
computation~\citep{London05}.

For our modeling purposes, we evoke the statistical physics principle
that emergent phenomena seem to depend on only few characteristics
such as symmetries, dimensionality, network topology, type of coupling
etc.  Hence the basic dynamical units (atoms, spins) need not be
modeled in detail~\citep{Marro99}. Our cellular automaton approach to
dendritic computation unveils the emergence of critical phenomena at
the single neuron level from a large number of sub-cellular
interacting units (dendritic compartments).

We find that the dendritic arbor can fire spontaneously when we take
into account the fact that dendritic spikes are non-stereotyped, that
is, with variable durations~\citep{Antic10}. In this case, a smooth
continuous phase transition appears between rest and self-sustained
dendritic activity.  At the interface between these states, there is a
critical regime, which optimizes the dynamic range of the input-output
response function of the neuron's dendritic arbor.  The presence of
the phase transition can increase the maximum neuronal signal
compression capacity by up to $\sim100$ times.  In our
\textit{critical-neuron} model, the minimal dynamic units are
dendritic branchlets (or perhaps patches of dendritic spines), so that
even a single neuron can have a highly nonlinear input-output response
function. This capability of a single neuron to compress stimulus
intensity varying over several orders of magnitude in a decade of
output firing rate could be the basis of psychophysical power
laws~\citep{Stevens75}.  In addition, such refinement in the basic
dynamical unit corresponds to a modeling improvement of spatial
resolution of a few orders of magnitude.

% Moreover, in consistency with the variability inherent of complex
% adapting biological systems, the analog dendritic computation we
% propose is comprehensive due to its robustness with respect to
% morphological and dynamical details of neurons.  Altogether, our
% results suggest that criticality constitutes a crucial factor also
% for building bioinspired and biomimetic sensors.

%\subsection*{Methods: The Model}
\section*{Results}
\label{sec:Results}
The overwhelming majority of models of criticality in neuronal
networks, if not all, neglect the characteristic tree shape of neurons. Typically,
the collective behavior studied is generated by oversimplified
point-like neurons with no spatial structure~\citep{BakChialvo03,
  Beggs04, Kinouchi06, Levina07, Rubinov11, Gal12, Gollo12b}.
Dropping this strong assumption unveils a new sublevel of neuronal
dynamics where concepts from statistical physics can be applied. In
previous work we introduced a simple model of active
dendrites~\citep{Gollo09}.  In the model, nonlinear excitable waves of
activity, not accounted for by passive dendrites and linear cable
theory, propagate and annihilate upon collisions, giving rise to large
signal compression abilities.  Weak inputs are highly amplified
whereas strong inputs are not subjected to early saturation. The
dynamic range, which quantifies how many decades of stimulus intensity
can be distinguished by the excitable medium, attains large values.
This property has proven robust against several model
variants~\citep{Gollo09, Gollo12}. 

In fact, excitable media with a tree structure performed better than
other network topologies~\citep{Assis08,Larremore11}. Moreover, the
system dynamics in a tree topology allows analytical treatment due to
the absence of loops.  The model was quantitatively understood by
solving the master equation under an excitable-wave (EW) mean-field
approximation~\citep{Gollo12}.  However, phase transitions are not
observed in such dendritic arbors when the dynamics of the excitable
elements (active dendritic branchlets) is deterministic.  In this
paper, we study a different dendritic arbor with non-stereotyped
dendritic spikes that exhibit probabilistic spike duration~\citep{Antic10},  
as well as non-homogeneous spike duration~\citep{Llinas80}, 
introduced by a layer-dependent function.  
Such biologically motivated improvements 
in the model give rise to a new dynamical regime consisting of a
self-sustained state with spontaneous activity that emerges through a
non-equilibrium phase transition.

Physiological experiments show that dendritic spikes depend on several
voltage-gated ionic channels.  Due to the non-homogeneous density of
channels along the proximal-distal axis, dendritic spikes can present
distinct dynamical features in different regions of the dendritic
tree.  Dominated by the slow calcium-dependent potentials~\citep{Llinas80,
  Stuart99,Schiller97} or N-methyl-D-aspartate (NMDA)
potentials~\citep{Antic10}, the dendritic spikes/plateaus have variable duration 
(see for example~Ref.~\citep{Antic10} and references therein), 
and the duration of active periods can become effectively longer at more distal sites~\citep{Llinas80}, 
as illustrated in Figs.~\ref{figDendrite}b through d.  
These responses account for a variety of active behaviors
including dendritic spikes with longer and more variable duration than 
sodium-dependent somatic spikes~\citep{Antic10}. 
%Na$^{+}$-dependent somatic spikes.

% \begin{figure}[!ht]
% \begin{center}
% \includegraphics[angle=0,width=0.99\columnwidth]{./hfigura1ModeloA.eps}
% \caption{\label{figDendrite} {\bf Model of an active dendritic tree 
%   with non-stereotyped dendritic spikes variable duration}. {\bf a},
%   Excitable elements (circles) connected (bars) in a Cayley tree
%   topology with $G=3$ layers and coordination number $z=3$ (one mother
%   and $k=2$ daughter branches).  {\bf b-d}, Biologically motivated 
%   dendritic spike with time duration depending on 
%   the distance from the soma as a net effect of variable density of ionic channels. 
%   Red arrows indicate pulses of external input. 
%   {\bf e}, Returning probability (see text for details) 
%   as a function of $p_\lambda$ (which governs the dendritic excitability) 
%   and $p_\delta$ (which governs the spike time duration).  
%   {\bf f}, Average firing rate as a function of $p_\lambda$ and $p_\delta$ 
%   for tree size $G=10$.}
% \end{center}
% \end{figure}

In our previous models~\citep{Gollo09, Gollo12}, other parameters have
been studied, but the time-length of the spikes that we introduce here
proves to be a crucial factor.  Such variability in the duration of
the spike %, which could be originated by voltage-gated membrane currents, 
has been shown to shape the network
dynamics~\citep{Manchanda13}, and to enhance the computational
capability of neuronal networks~\citep{Villacorta13}.  In the present
model, the variable duration of dendritic spikes plays an important
role in dendritic computation because it is capable of controlling the
emergence of a phase transition. Evidence for both sides of this
continuous non-equilibrium phase transition has been reported
experimentally~\citep{Llinas80,Johnston08}. Several different patterns of activity
observed experimentally are candidates for the active phase of the putative 
phase transition we propose: plateaus~\citep{Suzuki08,Antic10}, single
spikes~\citep{Davie08,Major08, Schiller97},
bursts~\citep{Wong79,Wong92}, and
oscillations~\citep{Kamondi98,Remme09}.

\vspace{0.25cm}

\paragraph*{\bf{Model.}} 
\label{sec:model}
As illustrated in Fig.~\ref{figDendrite}a, we model the dendritic
arbor as connected probabilistic cellular automata with a Cayley tree topology
of active branchlets capable of transmitting dendritic spikes. The
states update in parallel with discrete time steps of $\delta
t=1$~ms. Exploring the idea of a dendritic branchlet as a fundamental functional 
unit of the nervous system (see Ref.~\citep{Branco10} and references therein), 
we consider dendritic branchlets as excitable units. 
Each excitable unit (the $i$-th dendritic branchlet) has
three possible discrete states $s_i(t) \in \{0,1,2\}$.  A quiescent
site ($s_i(t)=0$) may become active in the next time step
($s_i(t+1)=1$) either by external driving or by propagation of an
excitable-wave from an active neighbor with probability
$p_\lambda$. External input arrives at quiescent sites with
probability $p_h=1-\exp(-h\delta t)$ per time step, where $h$ stands
for the rate of independent Poisson processes and may vary in a range
of several orders of magnitude. An active site ($s_i(t)=1$) becomes
refractory ($s_i(t+1)=2$) with probability $p^g_\delta$ at each time
step, where $g$ indicates the layer that the site belongs to (see
Fig.~\ref{figDendrite}c). To close the cycle, refractory sites become
quiescent with probability $p_\gamma$ (we fix $p_\gamma = 1/2$
throughout this paper).

The probabilistic nature of our model takes into account the fact 
that the density of ion channels in a given spatial region is not 
always large enough to allow their mean activation (or inactivation, 
or deactivation etc.) to be described by a continuous variable, 
and hence fluctuations around the mean are not negligible. 
Such non-deterministic and irregular factors have already been proven 
to play a crucial role in shaping the neuronal excitability~\citep{Carelli05,Cannon10}, 
the propagation~\citep{Jarsky05} and the duration of dendritic spikes~\citep{Antic10}.
Despite the fact that subthreshold dendritic activity can influence the
integration of synaptic signals~\citep{Remme09}, 
we assume a large attenuation of subthreshold activity to an extent 
that a boost of activity is required for activity propagation. 
It is noteworthy that our previous model~\citep{Gollo09}, which features this very same assumption, has been 
recently validated by detailed morphologically reconstructed multi-compartmental 
cell models with active dendrites but also including passive propagation of subthreshold activity~\citep{Publio12}. 

The model generalizes our previous work~\citep{Gollo09, Gollo12} by
introducing a probabilistic spike duration, 
which is motivated by experiments (see Ref.~\citep{Antic10} for a recent review paper).  
In what follows, we first consider a simple homogeneous model with spike duration
controlled by a uniform probability $p^g_\delta=p_\delta \leq1, \,
\forall \, g$ (our previous model~\citep{Gollo09} is thus recovered
when $p_\delta=1$).  For $p_\delta \simeq 0$ the spike duration
becomes unreasonably long.  Thus, our attention is mostly concentrated
in the regime between the extremes ($0<p_\delta<1$). 
Whereas the stereotyped sodium-dependent somatic spikes last typically 1 ms,
the duration of dendritic activity exhibits large variability
depending on the type of neuron and ion channels involved,
and fall within the range of 1-10 milliseconds (or even longer~\citep{Antic10}).
This is captured in our model by the fact that our time step $\delta t=1$~ms,
and the time spent in the active state is about  $p_\delta^{-1}$: For $p_\delta=1$,
the duration of our model dendritic spike is 1 ms, and for $p_\delta=0.1$
the average duration would be 10 ms.  
We start by analysing this model with homogeneous spatial distribution of spike duration, which has some advantages: 
it is clear, intuitive, and allows analytical insights about its phase transition, which is 
controlled by $p_\lambda$ and $p_\delta$.
 
Next, we study a more realistic scenario based on the relevant
physiological evidence previously described: the model assumes that
the spike duration depends on the proximal distance, 
such as previously reported in the Purkinje cell~\citep{Llinas80}.  
Owing to the absence of a clear functional dependence in the literature, 
for simplicity, we consider a
linear dependence $p^g_\delta=1-0.9 \frac{g}{G} \alpha$, where $G=10$
(unless otherwise stated) stands for the tree size, and the parameter
$\alpha$ ($0\leq \alpha \leq 1$) controls the inhomogeneity. 

%To characterize the dynamical regimes of our system, we must define our order parameters. 
%The usual one is the stationary density of active sites ~\citep{Marro99}:
%\begin{equation}
%\rho = \frac{1}{N}\sum_{i=1}^N \left< \delta (s_i(t),1) \right>_t\:, 
%\end{equation}
%where the temporal average is done in the stationary state and $\delta(a,b)$ is the Kronecker 
%delta, 
%it is one if $a=b$ and zero otherwise. 

To characterize the dynamical regimes of our system, we must define
our order parameter~\citep{Marro99}.  The main order parameter of
interest here is the stationary firing rate of the proximal site $s_0$
(see Fig.~\ref{figDendrite}a):
\begin{equation}
%F = \frac{\left< \delta (s_0(t),1)\right>_t }{\delta t}\:. 
%F = {\delta t}^{-1} {\left< \delta_{s_0(t),1}\right>_t }\:,
F = \frac{1}{T} \sum_{t=1}^T  \delta_{s_0(t),1}\:,
\end{equation}
where $T$ is an averaging time window and $\delta_{i,j}$ is the
Kronecker delta (by definition, $\delta_{i,j}=1$ if $i=j$, and zero otherwise).  The proximal dendrite
firing rate $F$ is biologically interesting because it provides the
main input to the neuronal soma and can be thought of as proportional
to the neuronal firing rate. Notice that $s_0$ and hence $F$ are
functions of the external input with intensity $h$.  One of our main foci
will be on the response function $F(h)$.

%\section*{Results}
%\label{sec:results}

\vspace{0.25cm}

\paragraph*{\bf{Phase diagram without external input.}} 
%\subsection{Phase diagram without external input.} 
\label{sec:spontaneousActivity}

It is instructive to analyze the complex spatio-temporal dynamics of
an extensive dendritic tree in the simplest scenario, when average
spike durations are homogeneous ($\alpha=0$) and the system is
subjected to no external driving ($h=0$). The absence of stimulus is a
somewhat artificial condition, but very informative because it crisply
uncovers the critical region. Starting with the states of nodes randomly 
distributed according to a uniform distribution among the three possible states, 
we follow the dynamics for a sufficiently long time ($T \sim
10^4$ time steps) until a stationary state is reached. If the tree
reaches an active and stable level, in which the activity in the dendritic tree does not vanish 
and the firing rate of the  proximal site is nonzero,  
the system is said to be supercritical. On the contrary, 
if the activity of the system vanishes rapidly, 
the system is subcritical. For large systems, the fate of the
system depends weakly on the initial condition and is mainly determined by
control parameters $p_\lambda$ (which controls the coupling between
branchlets) and $p_\delta$ (which controls the average spike
duration). The critical regime occurs at the border between the sub-
and supercritical regimes, where the activity level vanishes slowly
and without a characteristic time scale~\citep{Kinouchi06}.

\begin{figure}%[!ht]
\begin{center}
  \includegraphics[angle=0,width=0.85\columnwidth]{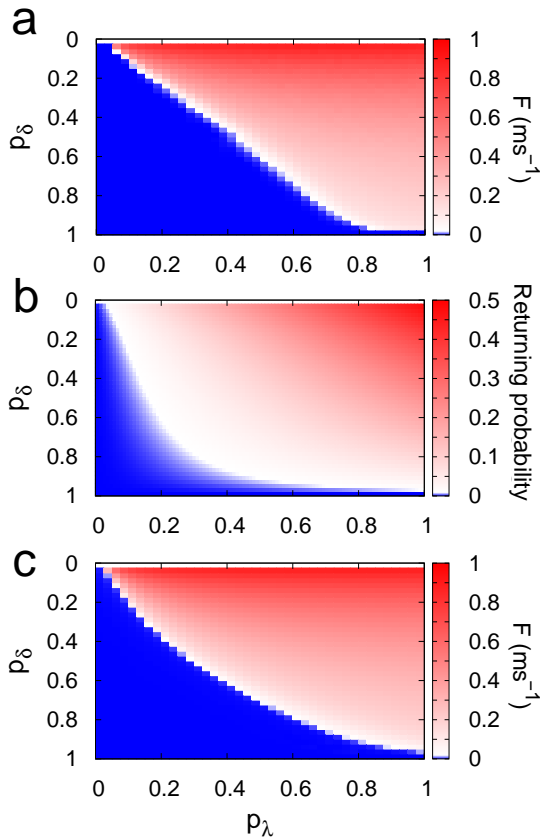}
  \caption{\label{figAAS} {\bf Continuous phase transition in active
      dendritic trees with non-stereotyped dendritic spikes}. {\bf (a)},
    Average firing rate as a function of $p_\lambda$ (which governs
    the coupling among branchlets) and $p_\delta$ (which governs the
    average spike duration) in numerical simulations for tree size
    $G=10$. {\bf (b)}, Returning probability (see Methods for details) as a
    function of $p_\lambda$ and $p_\delta$. {\bf (c)}, Same as panel {\bf (a)}
    but for the generalized excitable-wave mean-field approximation
    (see Methods for details).  }
\end{center}
\end{figure}

As represented in Fig.~\ref{figAAS}a, we numerically find the critical
curve of the parameter plane $(p_\lambda,p_\delta)$ in a finite tree
with $G=10$. The critical curve corresponds to the border between the
absorbing state with $F(h=0) = 0$ (blue) and the active state with
$F(h=0) > 0$ (red). Naturally, a true phase transition occurs only for
infinite systems, but for moderately large systems we already observe
activity that persists for simulation times much longer than
any relevant biological time scale. It is important to emphasize that
no active phase exists for dendritic spikes with a deterministic
duration, $p_ \delta=1$. Only when they are non-stereotyped ($p_\delta
< 1$) does self-sustained activity become possible.

%We now examine the phase diagram for an excitable tree that does not receive external input, that is,
%we determine the regions in the plane $p_\delta$~vs.~$p_\lambda$ where we have $\rho(h=0) > 
%0$ (or equivalently,
%where $F(h=0) > 0$. The results are presented in Figs.~\ref{figDendrite} and f.

To qualitatively understand why the activity dies out in a specific
part of the parameter space of a finite tree, we define the returning
probability $R$. It corresponds to the probability that an active site
$A$ stimulates a given quiescent neighbor $B$ and receives back the
stimulus at later times after performing a complete cycle (i.e.,
after going through the refractory ($s_A=2$) and the quiescent
($s_A=0$) states).  A necessary (but not sufficient) condition for a
network without loops to exhibit a stable self-sustained state (and
consequently a phase transition) is that $R\neq 0$. A nonzero returning probability is necessary for
the persistence of the self-sustained state because otherwise the
activity ceases after a few time steps due to collisions of the
excitable waves with the boundaries (layer $G$) and with one
another. The Methods
section\ref{sec:Methods} illustrates the fastest example of a
successful process of returning activity, which requires at least
three time steps, and extends the calculation for arbitrarily long
two-site processes. 
Figure~\ref{figAAS}b shows $R(p_\lambda,p_\delta)$, in which
a transition similar to that of Fig.~\ref{figAAS}a is observed. A
comparison between them confirms that $R\neq 0$ is a necessary but not
sufficient condition for an active phase to emerge.
 
%The returning probability $R$ as a function of the coupling
%($p_\lambda$) and the spike non-persistence ($p_\delta$) probabilities
%is shown in Fig.  Most of the parameter space region
%allows the activity to return (red).  However, low activity
%propagation probability, and more importantly, low spike
%non-persistence probability ($p_\delta \simeq 1$), forbids excitable
%waves to return to their origin (blue).
%
%
%Simulations of a finite system (see Fig.~\ref{figAAS}b) confirm the
%non-sufficiency condition of $R$ being nonzero for the self-sustained
%state to occur, since the stationary density of active sites is
%greater than zero in an area smaller than the area where $R > 0$,
%depicted in Fig.~\ref{figAAS}a.  This means that even in the presence
%of nonzero returning probability $R$, the self-sustained activity
%might turn out to be unstable for some finite systems.  

Another analytical approach to shed light on the results is to
employ one of the many mean-field approximations available in the
statistical physics literature~\citep{Marro99}. We have previously
developed an excitable-wave (EW) mean-field approximation that focuses
on the wave propagation direction, making it particularly suitable for
excitable waves on a tree~\citep{Gollo12}. In the Methods section we
describe a generalized version of the approximation to account for
non-stereotyped dendritic spikes. As Fig.~\ref{figAAS}c shows, the
phase transition observed in the simulations can be qualitatively
described by the generalized EW (GEW) mean-field approximation.

\vspace{0.25cm}

\paragraph*{\bf{Neuronal response to external input.}} 
%\subsection{Neuronal response to external input.}
\label{sec:externalfield}

From herein, we focus on the more relevant scenario in which a neuron
has to somehow cope with information arriving stochastically at its
thousands of synapses. We make the simplest assumption that, at each
branchlet, the average excess of incoming excitatory post-synaptic
potentials (EPSPs), as compared to inhibitory post-synaptic potentials
(IPSPs), can be modeled by an independent Poisson process with rate
$h$. We study the response function $F(h)$
(averaged over $T=10^4$ time steps and five realizations) and its
dependence on model parameters.

% From herein, we focus on the more relevant scenario in which a neuron
% has to somehow cope with information arriving stochastically at its
% thousands of synapses. We make the simplest assumption that, at each
% branchlet, the average excess of incoming excitatory post-synaptic
% potentials (EPSPs), as compared to inhibitory post-synaptic potentials
% (IPSPs), can be modeled by an independent Poisson process with rate
% $h$~\footnote{A variant of the model in which the rate $h$ varies
%   along the proximal-distal axis has been studied in
%   Ref.~\citep{Gollo09}.}. We study the response function $F(h)$
% (averaged over $T=10^4$ time steps and five realizations) and its
% dependence on model parameters.

\begin{figure*}%[!ht]
\centerline{\includegraphics[angle=0,width=1.7\columnwidth]{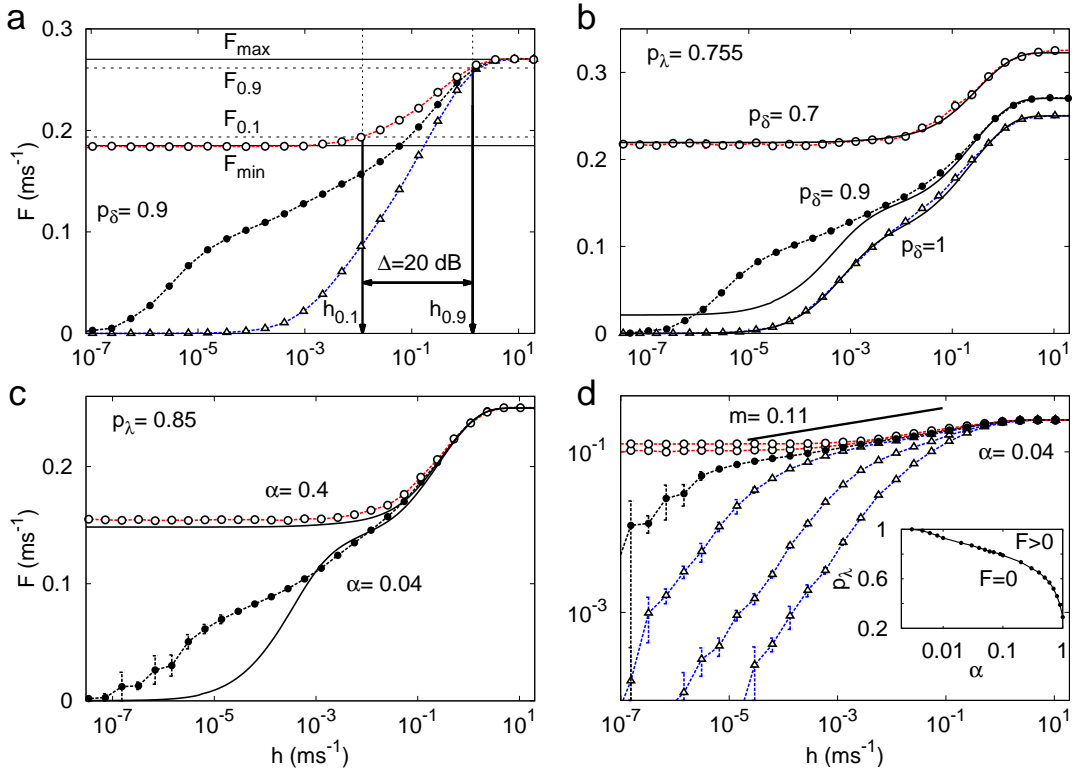}}
\caption{\label{figRF} {\bf Response functions and phase diagram}.
  {\bf (a)}, Response curves $F(h)$ for $p_\lambda= 0.5, 0.755, 1$ (from
  right to left).  Horizontal lines and vertical arrows show the relevant
  parameters for calculating the dynamic range $\Delta$ (see
  Eq.~(\ref{Delta})).  {\bf (b)}, Response curves dependence on $p_\delta$
  for homogeneous distribution $p^g_\delta=p_\delta \, \forall \, g$.
  Symbols connected by dashed lines represent simulation results
  whereas continuous lines represent the generalized excitable-wave
  (GEW) mean-field approximation.  {\bf (c)} and {\bf (d)}, Response
  curves dependence on $g$, i.e., $p^g_\delta=1-0.9 \frac{g}{G}
  \alpha$.  {\bf (d)}, Family of response curves for $p_\lambda= 0.4,
  0.6, 0.8, 0.85, 0.9, 1$ (from right to left). 
  Solid line represents a power law with exponent $m=0.11$, 
  which serves as a guide to the eye for the simulation results for the critical value $p_\lambda= 0.85$. 
  Inset: phase diagram. Tree size is $G=10$ for all panels.}
\end{figure*}

Figure~\ref{figRF}a depicts three response functions $F(h; p_\lambda)$
for a fixed $p_\delta$, exemplifying one response function of each
kind: subcritical (triangles, blue), critical (closed circles, black)
and supercritical (open circles, red).  Alternatively, for a fixed
$p_\lambda$, the critical line can also be crossed by varying
$p_\delta$, as depicted by Fig.~\ref{figRF}b.  In this case, the
maximum firing rate displays its dependence on $p_\delta$:
%$F_{max}=\frac{p_\delta^{-1}}{p_\delta^{-1}+3}$.
$F_{max}=({1+3 p_\delta})^{-1}$.

For some classes of dendrites, it can be more realistic to employ our
heterogeneous model, with $\alpha>0$. We consider $\alpha$ varying in
a logarithmic range of several decades because the phase transition
already occurs for $\alpha \ll 1$.  In this instance, the phase
transition is controlled by both $p_\lambda$ and $\alpha$, giving rise
to a critical curve in parameter space, as depicted in the inset of
Fig.~\ref{figRF}d.  Analogously to the homogeneous case, for fixed
$p_\lambda$ the transition depends on $\alpha$, as depicted in
Fig.~\ref{figRF}c.  For both homogeneous and heterogeneous models
(solid line of Figs.~\ref{figRF}b and c), the GEW mean-field
approximation captures particularly well the behavior of both sub- and
supercritical curves but fails to capture the strong amplification of
the critical curves under weak external driving. As shown in the
family of curves for fixed $\alpha$ displayed in Fig.~\ref{figRF}d,
the critical curve has a very small exponent ($m \simeq 0.11$), which
cannot be correctly described by mean-field approximations.  This
exponent $m$ is similar to a Stevens psychophysical response
exponent~\citep{Stevens75} observed at the single-neuron
scale~\citep{Gollo09}, as had been previously noticed at the level of
a neuronal ensemble~\citep{Copelli02, Furtado06, Kinouchi06}.

Excitable networks have a recognized ability to compress several
decades of input rate in a single decade of output
rate~\citep{Copelli02,Kinouchi06,Assis08,Larremore11,Gollo09, Gollo12,
  Gollo12b}.  This information processing capacity emerges exclusively
from local interactions between the excitable elements.  A large
dynamic range is robust, being also obtained for models with
increasing levels of biophysical realism, such as networks of
FitzHugh-Nagumo and Hodgkin-Huxley elements~\citep{Ribeiro08,
  Copelli02}, as well as more sophisticated models based on detailed
anatomical information of the retina~\citep{Publio09, Publio12}.

The definition of dynamic range is illustrated in
Fig.~\ref{figRF}a~\citep{Kinouchi06}.  We start by neglecting the
regions of the response curves which are close to the detection
threshold $F_{min}$ or the saturation level $F_{max}$.  
% Specifically,
% we cut out the regions within 10\% of these levels (an arbitrary but
% widely-used choice) because the $F(h)$ curve is barely invertible in
% these regions. 
According to a conventional definition, only the interval between
$F_{0.1}$ and $F_{0.9}$ (arbitrarily defined within 10$\%$ of the
plateau levels) properly codes the input.  Those firing rates
correspond to external driving with rates $h_{0.1}$ and $h_{0.9}$.
The dynamic range is the range (measured in decibels) between
$h_{0.1}$ and $h_{0.9}$, i.e.,
\begin{equation}
\label{Delta}
 \Delta \equiv10 \log_{10} \left(\frac{h_{0.9}}{h_{0.1}}\right).
\end{equation}

\begin{figure*}
\centerline{\includegraphics[angle=0,width=1.7\columnwidth]{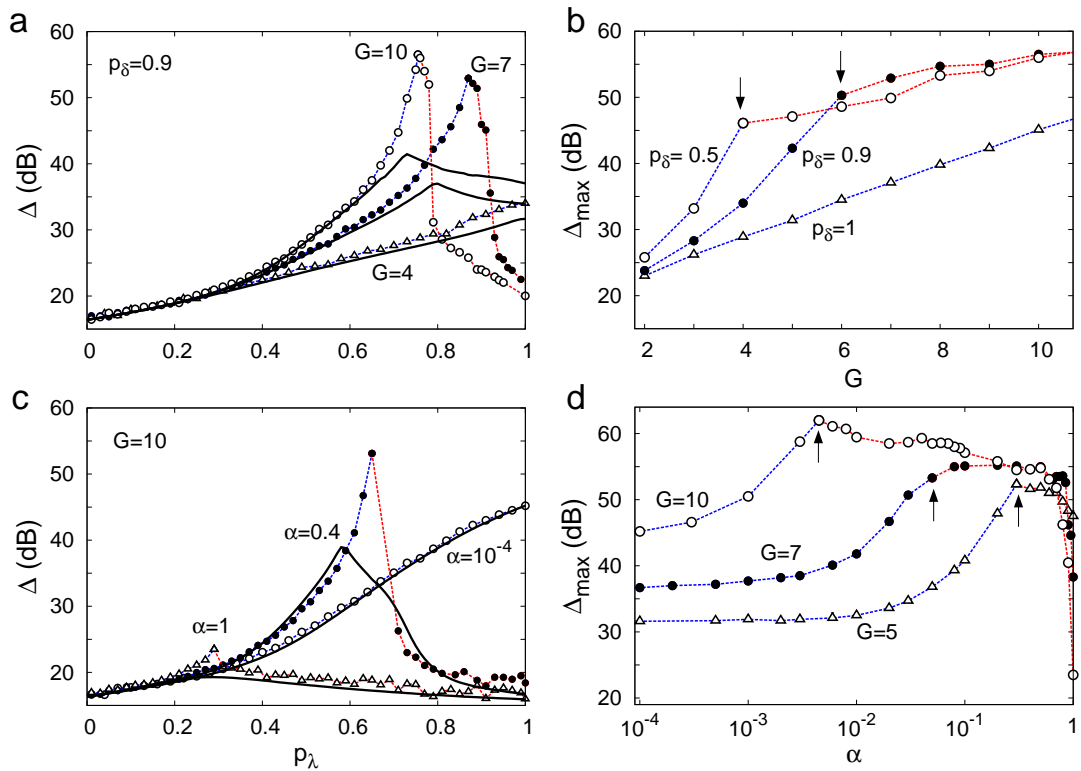}}
\caption{\label{figDelta} {\bf Dynamic range $\Delta$}.  {\bf (a)} and
  {\bf (b)}, for homogeneous distribution of $p_\delta$.  {\bf (c)} and
  {\bf (d)}, for layer-dependent distributions: $p^g_\delta=1-0.9
  \frac{g}{G} \alpha$.  {\bf (b)} and {\bf (d)} depicts $\Delta_{max}$
  which is the maximum of the $\Delta(p_\lambda)$ curve.  In each
  curve phase transitions occur only at and on the right hand side of
  arrows.  Symbols connected by dashed lines represent simulation
  results whereas continuous lines represent the generalized
  excitable-wave (GEW) mean-field approximation.}
\end{figure*}

The phase transition is an important feature for signal
compression. Comparing the different regimes, as depicted for instance
in Fig.~\ref{figRF}a, a simple reasoning allows one to understand why
the critical curve optimizes the compressing
capacity~\citep{Kinouchi06}.  The subcritical curves cannot amplify
small external stimuli, since the activity from incoming pulses
vanishes exponentially fast in space and time.  At the other extreme,
for supercritical curves, almost any pulse initiates the
self-sustained network activity for arbitrarily long time periods,
masking the response to weak stimuli.  However, when the network has
the correct critical parameters, the spontaneous network activity is
composed of neuronal avalanches launched by spontaneous fluctuations
of the system. In the presence of an external field $h$, such
avalanches add and superpose, creating the response function $F(h)$
(analogous to magnetization for non-zero fields in magnetic
systems). In this case, the network dynamic range is optimized.

%*****************************explicar cada caso sub/super
%**************************************************

Since phase transitions are properly defined only for infinite
systems, we have investigated the effects of system size in our
model. It turns out that, even for moderately large systems
(e.g. $G=10$, as in Figs.~\ref{figAAS}~and~\ref{figRF}),
self-sustained activity survives for a period of time which is much
longer than any relevant characteristic time for neuronal processing
($T\sim 10^4$~time steps $\sim$10 seconds), which is the operational
marker that we have employed for the transition. Its occurrence is
associated with the peaks in dynamic range shown in
Fig.~\ref{figDelta}a. 

In small trees, activity dies out rapidly due to wave collisions with
the boundaries, preventing self-sustained activity. Taking for example
the case of $G=4$ in Fig.~\ref{figDelta}a, the maximum $\Delta_{max}$
(of the $\Delta$ vs. $p_\lambda$ curve) occurs at $p_\lambda=1$,
meaning that finite size effects masks the (infinite size) phase
transition. Larger trees, in contrast, can exhibit
$\Delta_{max}=\Delta(p_{\lambda \, c})$ for weaker coupling revealing
the presence of a phase transition at $p_{\lambda \, c} (G)<1$. The
larger is the tree, the smaller is the $p_{\lambda \, c}(G)$
(Fig.~\ref{figDelta}a), and the larger is the $\Delta_{max}$
(Fig.~\ref{figDelta}b).

As presented in Fig.~\ref{figAAS}, $p_\delta=1$ prevents an active
state. However, for $p_\delta<1$ a phase transition to an active state
occurs in large enough trees. The black arrows in Fig.~\ref{figDelta}b
represent the minimum tree size that gives rise to our operationally
defined transition for $p_\delta=0.5$ ($G=4$), and $p_\delta=0.9$
($G=6$).

The optimization of the dynamic range at the critical value
(Fig.~\ref{figDelta}c), and the growth of the maximum dynamic range
with the tree size (Fig.~\ref{figDelta}d) are also observed in the
model with non-uniform spike duration.  In this more realistic model
version, the phase transition vanishes for $\alpha \sim 0$, as
depicted in the inset of Fig.~\ref{figRF}d, since in such case the spike
duration is deterministic across the tree.  Increasing $\alpha$ from
zero, $\Delta_{max}(\alpha)$ grows up to a maximum around the region
where the phase transition appears (represented by arrows in
Fig.~\ref{figDelta}d).  As shown in Fig.~\ref{figDelta}c, larger
values of $\alpha$ lead to a phase transition with smaller critical
coupling $p_{\lambda \, c}$.  On the other hand, as depicted in
Fig.~\ref{figDelta}d, the $\Delta_{max}(\alpha)$ curves show plateaus
with heights and widths that both increase with tree size.  This
suggests that large active dendritic trees with non-uniform spike
duration lead to a large dynamic range, a result whose robustness is
attested by the width of the plateaus.  In particular, the difference
in dynamic range for an optimized $\alpha$ compared with the
$\alpha=0$ case is remarkable, attaining about $20$ dB for $G=5$
(Fig.~\ref{figDelta}d).

% Finally, we notice that the GEW approximation qualitatively captures
% the dynamic range behavior of our model.  However, it is not able to
% capture the strong amplification of the response functions at
% criticality for weak external driving, see Figs.~\ref{figRF}b and c,
% missing the dynamic range starting point $h_{0.1}$.  Therefore, as
% shown in Figs.~\ref{figDelta}a and d, in the presence of a phase
% transition, the dynamic range under the GEW approximation is reduced
% around the critical region $p_{\lambda \, c}$.

\section*{Discussion}
\label{sec:discussion}

%\subsection*{Conclusion}
%\label{sec:conclusion}

We have shown that the idea of a critical network of excitable
elements being able to optimally process incoming stimuli can be
applied at the subcellular (dendritic) scale. Instead of a network of
excitable point-like neurons connected by synapses, here we have shown
that electrically connected excitable dendritic branchlets can cause a
dendritic arbor of a single neuron to undergo a phase transition. The
key ingredient for this scenario is stochasticity in the duration of
dendritic spikes.

Several experiments indicate that a dendritic spike at a distal site
is usually not strong enough to trigger a somatic action
potential~\citep{Jarsky05}. This suggests that real dendrites ought
not typically show deterministic signal propagation (that is,
$p_\lambda < 1$). In the present model, taking for instance a dendrite
with tree size $G=10$ and probability of activity propagation
$p_{\lambda}=0.5$, the probability for a most distal dendritic spike
to propagate all over to the proximal site and potentially generate a
spike is very small: $(0.5)^{10}\simeq10^{-3}$.  Thus, the model is in
accordance with experimental wisdom. Remarkably, whenever the
dendritic parameters lie near the silent/active phase transition, our
results prove that non-reliable dendritic spike propagation is
compatible with an optimal dynamic range.

% This work is a first step in the quest for the elementary dynamical
% unit for neuronal computation.  We show that such elementary unit
% could be, at least, as small as a dendritic branchlet (or perhaps even
% a dendritic spine).  The elementary unit in a critical system is
% seldom well defined, but at least the frontiers can become more
% accurately set.

We have examined the implications of variable dendritic spikes (non-homogeneous
and non-deterministic) for dendritic computation.  We
have shown that a necessary condition for a phase transition to occur
in active dendrites is that the returning probability $R$ be nonzero,
which in turn can only occur with variable dendritic
duration ($p_\delta<1$). In an attempt to incorporate more details on
dendritic spikes, we have also modeled different spatial functions of
average spike duration, and showed that our main results are
robust in that parameter space.

While the variable-duration criterion must be satisfied to give rise to
a critical neuron, it does not play such a crucial role in giving rise
to a critical neuronal network~\citep{Manchanda13}. The key difference
between the two spatial scales is the topology: due to inhibitory
signaling mechanisms during growth~\citep{Jan10}, a dendritic tree
contains no loops, whereas in neuronal networks they abound. The
possibility of criticality at both scales naturally raises the
question: if neurons can be critical, why should we need critical
networks? We propose that the answer depends on the type of neuronal
computation and the scale at which it occurs. In the cases where the
processing is distributed among a large number of neurons with small
dendritic trees, it is plausible to look at point-like neurons as the
basic units of a larger system which, via balanced synapses, can tune
itself to a collective critical regime. By contrast, single-neuron
criticality could play a computational role in cases where neurons
have extended dendritic trees and therefore must cope with large
variations of synaptic input (such as olfactory mitral cells or
cerebellar Purkinje cells, for instance). Clearly, in principle
nothing prevents the mechanisms at both scales from acting together.

The input-output response function of a critical neuron amplifies
small-intensity inputs while preventing early saturation when
subjected to large-intensity inputs.  The response function follows
a slow-increasing power-law function $F\sim h^{m}$, in which $m$
corresponds to a rather tiny exponent $ m \simeq 0.11$, as shown in
Fig.~\ref{figRF}d.  Such a slow-increasing function could be easily
confounded with a logarithmic (Weber-Fechner's psychophysical)
law~\citep{Stevens75,Copelli02}.  
% Moreover, the exponent seems
% compatible with the direct percolation universality class in dimension
% $d=1$~\citep{Marro99}.  Very small psychophysical Stevens' exponents
% are found in visual, somatosensory, olfactory, gustatory systems among
% others~\citep{Stevens75}.
These results strengthen previous suggestions that a system poised at
a critical point is a natural candidate to explain how $m<1$ Stevens'
power-law exponents emerge~\citep{Kinouchi06}.

% In the rate-code framework, it is fundamental for neurons and neuronal
% systems to exhibit an amplification of small inputs without early
% saturation.  Response functions with these elements exhibit a large
% dynamic range. In fact, the larger is the neuronal dynamic range, the
% greater is the neuronal capacity to code, and presumably the capacity
% of the animal to survive. 

Our simple model yields similar results to those obtained with
detailed compartmental modeling of morphologically reconstructed
dendrites: the maximum dynamic range a dendritic arbor can attain
grows with the tree size~\citep{Publio12}. This supports the previous
proposal that neurons with large dendritic arbors might have grown to
attain such impressive size and complexity in order to enhance their
capacity to distinguish the amount of external driving~\citep{Gollo09,
  Gollo12}.
 
% Our results extend to a smaller scale the widely accepted fact that a
% critical system optimizes the dynamic range~\citep{Kinouchi06,
%   Assis08, Shew09,Larremore11,Gollo12b}, in particular confirming that
% the maximum dynamic range a dendritic arbor can attain grows with the
% tree size~\citep{Gollo09, Gollo12, Publio12}. This supports the
% previous proposal that neurons with large dendritic arbors might have
% grown to attain such impressive size and complexity in order to
% enhance their capacity to distinguish the amount of external
% driving~\citep{Gollo09, Gollo12}.  
% On top of that, our more detailed model also shows that large trees
% give rise to maximum dynamic ranges with wide and tall plateaus,
% which make them less susceptible to a precise fine tuning of the
% parameters.  Such robustness is crucial in neuronal systems that are
% subjected to a multitude of overlapping dynamical behaviors:
% plasticity, growth, pruning, depression, facilitation, adaptation,
% and homeostasis.

Owing to our irresistible tendency of comparing brain processes with
whichever technology happens to be dominant at the time, the term
dendritic computation is typically associated with a symbolic and
digital-like information processing~\citep{Koch}.  However,
%digital-like information processing~\citep{Koch,Caze13}.  However,
dendritic trees are living tissues that change all the time, growing
and retracting branchlets, spines and synapses. For example, 30\% of
spine surface retracts in hippocampal neurons over the rat estrous
cycle~\citep{Woolley90}. This, we believe, does not seem compatible
with a fixed circuitry implementing, say, logical gates. In
contrast, our framework suggests that dendritic arbors perform a
robust \textit{analog} computation, which is resilient to disturbances
of tree properties. For instance, pruning half of the tree corresponds
to changing from $G$ to $G-1$, which amounts to a small decrease in
the dynamic range.

% releases the system from such unreasonable constraints,
% such as boolean logic-gates at the dendritic level. Our simple model
% of dendritic computation equipped with non-stereotyped dendritic
% spikes shows that dendritic arbors perform a robust dynamical and
% \textit{analog} computation consistent with the current knowledge of
% the physiology of neuronal systems, and which is plausibly resilient
% with respect to disturbances of the fine-integrity of neuronal
% circuits.

In contrast to physical systems, biological systems can approach the
critical state through homeostatic mechanisms~\citep{Levina07,
  Bonachela10}. This self-organized tuning is essential for processing
information over a large range of stimulus intensities.  We believe that
critical neurons do indeed exist, but experimental limitations or lack
of a clear theoretical framework possibly prevented spatio-temporal
neuronal criticality from having been reported in the literature to date. 
It is important to emphasize, however, that our model focusses on a regime dominated by  
the active propagation of activity (dendritic spikes). Whether or not a phase transition occurs in a regime where active as well as passive mechanisms coexist remains untested and should be investigated 
in future models and experiments.

% In our critical
% neuron model, the minimal dynamic units are the dendritic branchlets
% (or perhaps patches of dendritic spines), so that even a single neuron
% can have a highly nonlinear input-output ``psychophysical neuronal
% law'' ~\citep{Stevens75}. Such refinement in the basic dynamical unit
% corresponds to an improvement of spatial resolution of a few orders of
% magnitude.

Experimental confirmation of single-neuron criticality requires
spatial and temporal recording resolutions at the edge of current
available techniques. %, which are likely to be available in the near future. 
Several lines of evidence already suggest critical phenomena
at the neuron level. {\it In vivo} long-range inter-spike time
interval correlations, which may have originated from a critical
neuron, have been reported in human neurons from
hippocampus~\citep{Parish04} and amygdala~\citep{Bhattacharya05}, as
well as in rat neurons from the leech ganglia and the
hippocampus~\citep{Mazzoni07}.  Furthermore, neglecting the extensive
spatial features of neurons, criticality in neuronal excitability has
recently been proposed~\citep{Gal12}.

Signs of a self-sustained state could be associated to patterns of
dendritic activation: spikes, plateaus, bursts, and
oscillations~\citep{Llinas80}. Evidently, dendritic trees can also be
on the other side of the transition, showing a silent or rest state.
Our model predicts a continuous phase transition so that criticality
lies between such subcritical (silent) and supercritical
(self-sustained active) regimes. We speculate that adaptation and
homeostatic mechanisms, similar to those discussed in
Refs.~\citep{Levina07, Bonachela10}, could finely tune a
self-organized critical state (or more precisely, a self-organized
quasi-critical state~\citep{Bonachela10}).  Such homeostatic
mechanisms, conceivably present at the dendritic level, will be
presented in a forthcoming paper.

%%%%%%%%%%%%%%

% Comparar aqui nossas propostas com outras propostas de computacao
% dendritica na literatura. Marcar territorio afirmando que nossa
% proposta eh de uma computacao analogica feita a partir de elementos
% e estruturas biologically non-reliable: mesmo deletando galhos da
% arvore, o range dinamico continua sendo funcao basicamente do numero
% N de sitios conectados. Isso contrasta com todas as propostas
% anteriores onde a computacao dendritica eh vista como digital-like e
% dependente da manutencao de uma integridade fina dos circuitos
% neuronais.

% The phase transition can be captured by GEW approximation, however, it
% is only qualitatively, specially failing to capture the correct
% critical behavior.  It might be necessary to go beyond a mean-field
% approximation to correctly describe such behavior.  Notice, however,
% that this phase transition only appears in models with a huge number
% or compartments or branchlets and are invisible in traditional models
% with moderate (some dozens) of compartments~\citep{Koch}.  These are
% important aspects that must be taken into account for designing, say,
% artificial stimulus detectors~\citep{Medeiros12}.

On the computational side, our cellular automaton model can be
generalized so that each branchlet is modeled by a detailed
biophysical compartment with a plethora of ion
channels~\citep{Publio12}, coupled with (noisy) axial
resistances. However, as in the proverbial arbor that prevents us from
seeing the forest, we must be aware that detailed biophysical modeling
may hamper us from detecting phase transitions if the model has few
compartments. Our statistical physics model, with simple excitable
units but connected in a large dendritic network, enables us to see
the forest.

\section{methods}
\label{sec:Methods}

\vspace{0.25cm}

\paragraph*{\bf{Returning probability calculation. }} 

\begin{figure}%[!ht]
\centerline{\includegraphics[angle=0,width=0.59\columnwidth]{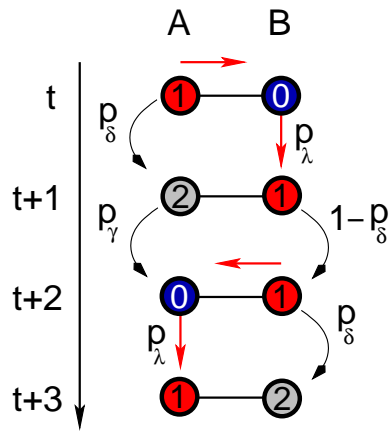}}
\caption{\label{figPreturn} {\bf Example of returning activity}.  An
  active site A stimulates its neighbor B (red arrow) and receives
  back the activity three time steps later. This process involves a
  series of intermediate steps.  Site A must pass through the
  refractory ($s_A(t+1)=2$) and quiescent ($s_A(t+2)=0$) states before
  becoming active ($s_A(t+3)=1$) again.  Moreover, site B must be kept
  in the active state ($s_B(t+1)=1$) for at least one time step
  ($s_B(t+2)=1$) in order to be able to excite back the susceptible A
  site ($s_A(t+2)=0$).  Therefore a nonzero persistence probability
  ($1-p_\delta > 0$) is fundamental for a nonzero returning
  probability ($R>0$).  }
\end{figure}

%\subsection{\label{Preturn} Returning probability calculation.} 
The returning probability $R>0$ is a necessary condition for a phase
transition.  For the phase transition between quiescent and
self-sustained states to occur the system must allow the
self-sustained activity to be kept for arbitrarily long times.  In a
finite network without loops, such as a Cayley tree, this condition is
only satisfied if the activity can go back and forth between two
neighbor sites, say $A$ and $B$, as illustrated in
Fig.~\ref{figPreturn}.  Otherwise, in the absence of external driving,
the system must go to the rest state (regardless of the initial
condition) after a maximum of $2G+1$ time steps.

It is possible to calculate the returning probability between two
neighbor sites that obey the cyclic cellular automata rules. The
example of Fig.~\ref{figPreturn} can be computed directly. Considering
the initial condition at time $t$, $s_A(t)=1$ and $s_B(t)=0$, there is
a unique possible path for the excitable wave to go back and forth
after three time steps.  Notice that, by construction, $s_A(t)$ must
complete the cycle (i.e., it must go through states $s_A=2$ and
$s_A=0$), since we have excluded the trivial solution: $s_A(t)=1,
\forall t$.  The final configuration $s_A(t+3)=1$ and $s_B(t+3)=2$ is
therefore reached with probability $p_\delta^2 (1-p_\delta)
p_\lambda^2 p_\gamma$, where $p_\delta p_\gamma p_\lambda$ is the
probability of site A to go throughout the cycle, and $p_\lambda
(1-p_\delta) p_\delta$ is the probability of site B to receive the
input, to remain active, and finally to become refractory.

This example is the simplest one. In general, however, there are
infinitely many possibilities that must be included in a full
calculation of $R$.  The configuration $s_A=2$ and $s_B=1$, which was
depicted at time $t+1$ in Fig.~\ref{figPreturn}, could have also be
found at later times: $t+2$, $t+3$ and so on, leading to identical
final configuration $s_A=1$ and $s_B=2$. The calculation of the
returning probability which accounts for all the possible intermediate
steps, considering that site $B$ has just been activated
($s_B(t+1)=1$), involves three sums of infinite geometric series,
$S_1$, $S_2$ and $S_3$.  Site $A$ can spend arbitrarily long periods
active ($\propto S_1$), refractory ($\propto S_2$), or quiescent
($\propto S_3$) and nevertheless receive back the excitable wave,
given that the pivot site $B$ remains active in the meantime. Finally,
without loss of generality, we can ignore the final state of site $B$
to find  the returning probability after infinitely many time steps:
\begin{eqnarray}
\label{ReturningP} R &=&  p_\delta p_\gamma (1-p_\delta) p_\lambda^2 S_1 S_2 S_3 \;,
\end{eqnarray}
where
\begin{eqnarray}
\label{s1} S_1 &=&  \frac{1}{1-(1-p_\delta)^2} \;, \\
\label{s2} S_2 &=&  \frac{1}{1-(1-p_\gamma)(1-p_\delta)} \;, \\
\label{s3} S_3 &=&  \frac{1}{1-(1-p_\lambda)(1-p_\delta)} \;.  
\end{eqnarray}
%\begin{eqnarray}
%\label{ReturningP} R &=&  \frac{p_\delta p_\gamma (1-p_\delta) p_\lambda^2}
%{[1-(1-p_\delta)^2][1-(1-p_\delta)(1-p_\gamma)][1-(1-p_\delta)(1-p_\lambda)]}  .
%\end{eqnarray}

Equations~(\ref{ReturningP})-(\ref{s3}) were used in
Fig.~\ref{figAAS}b.  However, this result can also be extended
to account for the heterogeneous case with $p_\delta$ belonging to
different layers ($a$ and $b$).  The returning probability $R^{a,b}$
turns out to be
\begin{eqnarray}
\label{ReturningPab} R^{a,b} &=&  p_\delta^a p_\gamma (1-p_\delta^b) p_\lambda^2\, S_1^{a,b}\, 
S_2^{a,b}\, S_3^{a,b}\, \; ,
\end{eqnarray}
where
\begin{eqnarray}
\label{s1ab} S_1^{a,b} &=&  \frac{1}{1-(1-p_\delta^a)(1-p_\delta^b)} \;, \\
\label{s2ab} S_2^{a,b} &=&  \frac{1}{1-(1-p_\gamma)(1-p_\delta^b)} \;, \\
\label{s3ab} S_3^{a,b} &=&  \frac{1}{1-(1-p_\lambda)(1-p_\delta^b)} \;.  
\end{eqnarray}

\vspace{0.25cm}

\paragraph*{\label{GEW} \bf{Generalized excitable-wave mean-field approximation.}} %\bf{Returning probability calculation. }} 

%\subsection{\label{GEW} Generalized excitable-wave mean-field approximation.} 
This approximation generalizes the recently proposed
excitable-wave (EW) mean-field calculation~\citep{Gollo12}.  Here, the
scope of the approximation is enlarged to account for non-stereotyped
dendritic spikes with probabilistic spike duration ($p_\delta
\leq 1$) and non-homogeneous spatial distributions ($p_\delta(g)$).
This is a single-site mean-field approximation, but it keeps track of
the excitable-wave direction of propagation.  Remarkably, the system
response is better captured by this approximation than by the
traditional two-site mean-field approximation~\citep{Gollo12}.

First we explain the notation and then recall the system master
equations.  Assuming that at time $t$ a site at generation $g$ is in
state $x$; its mother site at generation $g-1$ is in state $y$; and
$i$ ($j$) of its daughter branches at generation $g+1$ are in state
$z$ ($w$) etc., the joint probability of this configuration reads as:
$P_t^g\left(y;x;z^{(i)},w^{(j)},\ldots\right)$.  We also employ the
usual normalization conditions $\sum_{w_3}
P_t^g\left(y;x;w_1,w_2,w_3\right) = P_t^g\left(x;y;w_1,w_2\right)$.
Thus, the master equations for arbitrary coordination
number $z=k+1$ and layers $0<g<G$ is given by~\citep{Gollo12}:
\begin{eqnarray}
\label{P(1)}
P_{t+1}^{g}(;1;) & = & P_{t}^{g}(;0;)  - \left(1-
p_h \right) \nonumber \\
& & \cdot    \sum^{k}_{i=0}\biggl[   p_{\lambda}^i {k
\choose i} (-1)^i P_t^{g}\left(;0;1^{(i)}\right) \nonumber \\
 & & \nonumber \\
 & &- \beta p^{i+1}_\lambda{k \choose i}(-1)^i
P_t^{g}\left(1;0;1^{(i)}\right)\biggr] \nonumber \\
 & & +(1-p_\delta) P_t^{g}(;1;)\;,  \\
\label{P(2)}
P_{t+1}^{g}(;2;) & = & p_\delta P_{t}^{g}(;1;)  + (1-p_\gamma) P_{t}^{g}(;2;)\; , \\
\label{P(0)}
P_{t+1}^{g}(;0;)& = & 1 - P_{t+1}^{g}(;1;) - P_{t+1}^{g}(;2;)\;,
\end{eqnarray}
where $P_t^g(y;x;w^{(0)})\equiv P_t^g(y;x;)$ is a two-site joint
probability. For simplicity, we also drop the excess of notation
symbols ($;$), such that $P_t^g(x)$ stands for $P_t^g(;x;)$ which is
the probability of finding at time $t$ a site at generation $g$ in
state $x$ (regardless of its neighbors).

Equations controlling the active state for sites belonging to the
first ($g=0$) and last ($g=G$) layer can be obtained from
straightforward modifications of Eq.~(\ref{P(1)}), yielding:
\begin{eqnarray}
\label{P0(1)}
P_{t+1}^{0}(;1;) & = & P_{t}^{0}(;0;)  - \left(1-
p_h \right)  \nonumber \\
& & \cdot \sum^{k+1}_{i=0}\biggl[   p_{\lambda}^i {k+1
\choose i} (-1)^i P_t^{0}\left(;0;1^{(i)}\right)\biggr] \nonumber \\
 & & +(1-p_\delta) P_t^{0}(;1;)\;,  \\
\label{PG(1)}
P_{t+1}^{G}(;1;) & = & P_{t}^{G}(;0;) + (1-p_h)\left[\beta p_\lambda P^{G}_t(1;0;)\right]  \nonumber 
\\
& &   +(1-p_\delta) P_t^{G}(;1;)\;, 
\end{eqnarray}
while equations controlling the refractory~(\ref{P(2)}) and
quiescent (\ref{P(0)}) states remain unchanged. The full description
of the dynamics requires higher-order terms (infinitely many in the
limit $G\to\infty$), but, as we show below,
Eqs.~(\ref{P(1)})-(\ref{P(0)}) are enough for the GEW approximation.

\begin{figure}%[!ht]
%\centerline{\includegraphics[angle=0,width=0.69\columnwidth]{Fig6.eps}}
\centerline{\includegraphics[angle=0,width=0.99\columnwidth]{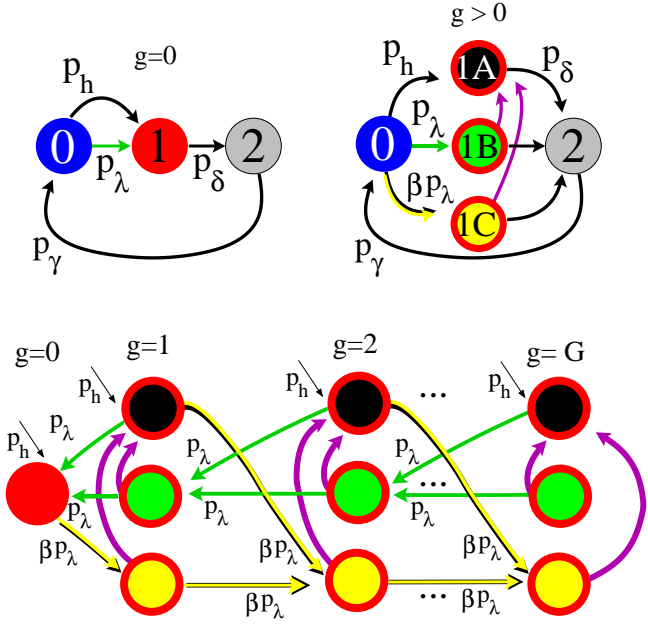}}
\caption{\label{GEWMFA} {\bf Generalized excitable-wave mean-field
    approximation}.  Top-left panel illustrates the dynamics of each
  node and of layer $g=0$ under the GEW mean-field approximation.
  Top-right panel illustrates the dynamics of further layers under the
  GEW mean-field approximation.  Bottom panel shows the
  excitable-waves direction of propagation.  Notice that in the absence of 
  purple arrows the flux vanishes in the state $(1C)$ of the last
  layer.  The purple arrows represent transitions that occur at each time step with
  probability $(1-p_\delta)^2$ (see text for details).  This
  generalization allows a qualitative description of the system
  for $p_\delta<1$.  }
\end{figure}

The rationale for the EW approximation (restricted to $p_\delta=1$) is
simple~\citep{Gollo12}: in an excitable tree, activity is decomposed
in forward- and backward-propagating excitable waves. We separate (for
$g>0$) the active state $(1)$ into three different active states:
$(1A)$, $(1B)$, and $(1C)$, as represented in
Fig.~\ref{GEWMFA}. $P_t^g(1A)$ (black) corresponds to the density of
active sites which received an external input and thus generates
excitable-wave propagating both forwards and backwards.  $P_t^g(1B)$
(green) corresponds to the density of active sites which received only
forward-propagating input.  Finally, $P_t^g(1C)$ (yellow) corresponds
to the density of active sites which received only backward-propagating
input.

For a correct description of the general case, $p_\delta \leq 1$, we
need to introduce loops into the topology as represented by the purple
arrows of Fig.~\ref{GEWMFA}.  First, we assume that with probability
$1-p_\delta$ an active site remains active for the next time
step. Second, among those sites that remained active we consider that
a fraction $1-p_\delta$ of them can jump to $P_{t+1}^g(1A)$, and
therefore contribute with both forward- and backward-propagating
excitable-waves.  This last step mimics the actions of the
returning activity, since state $(1A)$ is a required intermediate step
to fulfill a back and forth movement of the excitable-wave under the
GEW mean-field approximation.

Following these ideas, and applying the usual mean-field
approximations~\citep{Gollo12}, one can write the equations for the
$g>0$ layers as
\begin{eqnarray}
\label{P1A}P^g_{t+1}(1A)&=& P^g_t(0) \Lambda^g_A + (1-p^g_\delta)  \nonumber \\
 & & \cdot \bigl\{ P^g_t(1A)+ (1-p^g_\delta)  \nonumber \\ 
& & \cdot \bigl[P^g_t(1B)+P^g_t(1C)\bigr] \bigr\} \; , \\ 
\label{P1B}P^g_{t+1}(1B)&=& P^g_t(0) (1-\Lambda^g_A) \Lambda^g_B(t) \nonumber \\
& &  + p^g_\delta (1-p^g_\delta) P^g_t(1B)  \; , \\ 
\label{P1C}P^g_{t+1}(1C)&=& P^g_t(0) (1-\Lambda^g_A)  \left[1-\Lambda^g_B(t)\right]
\Lambda^g_C(t)  \nonumber \\
& &  + p^g_\delta (1-p^g_\delta) P^g_t(1C)\; ,
\end{eqnarray} 
where the excitation probabilities are given by
\begin{eqnarray}
\label{LambdaA}\Lambda^g_A &=& p_h,  \\
\label{LambdaB}\Lambda^g_{B} (t)&=& 1-\bigl\lbrace 1-p_\lambda \bigl[P^{g+1}_t(1A) \nonumber \\
 & &  +P^{g+1}_t(1B)\bigr] \bigr\rbrace^k, \\
\label{LambdaC}
\Lambda^g_{C} (t)&=& \beta p_\lambda \left[P^{g-1}_t(1A)+P^{g-1}_t(1C)\right].
\end{eqnarray}

Equations~(\ref{P(2)})~and~(\ref{P(0)}) remain unchanged, with $
P^{g}_{t}(1) \equiv P^g_{t}(1A)+P^g_{t}(1B)+P^g_{t}(1C)$. The dynamics
of the most distal layer $g=G$ is obtained by fixing
$\Lambda_B^{G}(t)=0$. The proximal element ($g=0$) has a simpler
dynamics since it does not receive backpropagating waves, so its
activity is simply governed by
\begin{eqnarray}
\label{P1g0}P^0_{t+1}(1)&=& P^0_t(0) \Lambda^0(t) + (1-p^0_\delta) P^g_t(1)  \; ,  
\end{eqnarray}
with
\begin{eqnarray}
 \Lambda^0(t) & = &
1-(1-p_h)\left\lbrace 1-p_\lambda
  \left[P^{1}_t(1A)+P^{1}_t(1B)\right]\right\rbrace^{k+1}. \nonumber 
\end{eqnarray}

Taking into account the normalization conditions, the dimensionality
of the map resulting from the GEW approximation is the same as for the
EW approximation~\citep{Gollo12}: $4(G-1)+5$. Numerical solutions of
this map with $\beta=1$ leads to the solid curves in Fig.~\ref{figRF} and
\ref{figDelta}.

%\end{methods}

\vspace{0.25cm}

%% Here is the endmatter stuff: Supplementary Info, etc.
%% Use \item's to separate, default label is "Acknowledgements"
\begin{acknowledgments}
   We are thankful to James A. Roberts for a careful reading of the manuscript. 
  The authors acknowledge financial support from Brazilian agency
  CNPq. LLG and MC have also been supported by FACEPE, CAPES and
  special programs PRONEX.  
  MC has received financial support from PRONEM, and OK and MC have been supported by CNAIPS-USP.  This research was also supported by a grant from MEC
  (Spain) and FEDER under project FIS2007-60327 (FISICOS) to LLG.
\end{acknowledgments}

% 
%  \item[Competing Interests] The authors declare that they have no
% competing financial interests.
%  \item[Correspondence] Correspondence and requests for materials
% should be addressed to L.L.G. \\ (email: leonardo.l.gollo@gmail.com).
% 
% \item[Author Contributions]  Conceived and designed the simulations: 
% LLG, OK and MC. Performed the simulations and analyzed results: 
% LLG. Contributed to analysis: OK and MC. Wrote the paper: LLG, OK and MC.

%\bibliography{./2011GolloCopelli}

\begin{thebibliography}{10}
\expandafter\ifx\csname url\endcsname\relax
  \def\url#1{\texttt{#1}}\fi
\expandafter\ifx\csname urlprefix\endcsname\relax\def\urlprefix{URL }\fi
\providecommand{\bibinfo}[2]{#2}
\providecommand{\eprint}[2][]{\url{#2}}

\vspace{0.5cm}

\bibitem{Chialvo10}
\bibinfo{author}{Chialvo, D.~R.}
\newblock \bibinfo{title}{Emergent complex neural dynamics}.
\newblock \emph{\bibinfo{journal}{Nat Phys}} \textbf{\bibinfo{volume}{6}},
  \bibinfo{pages}{744--750} (\bibinfo{year}{2010}).

\bibitem{Sornette00}
\bibinfo{author}{Sornette, D.}
\newblock \emph{\bibinfo{title}{Critical Phenomena in Natural Sciences}}
  (\bibinfo{publisher}{Springer}, \bibinfo{address}{Berlin},
  \bibinfo{year}{2000}).

\bibitem{Beggs03}
\bibinfo{author}{Beggs, J.~M.} \& \bibinfo{author}{Plenz, D.}
\newblock \bibinfo{title}{Neuronal avalanches in neocortical circuits}.
\newblock \emph{\bibinfo{journal}{J Neurosci}} \textbf{\bibinfo{volume}{23}},
  \bibinfo{pages}{11167--11177} (\bibinfo{year}{2003}).

\bibitem{Beggs04}
\bibinfo{author}{Beggs, J.~M.} \& \bibinfo{author}{Plenz, D.}
\newblock \bibinfo{title}{Neuronal avalanches are diverse and precise activity
  patterns that are stable for many hours in cortical slice cultures}.
\newblock \emph{\bibinfo{journal}{J Neurosci}} \textbf{\bibinfo{volume}{24}},
  \bibinfo{pages}{5216--5229} (\bibinfo{year}{2004}).

\bibitem{Friedman12}
\bibinfo{author}{Friedman, N.} \emph{et~al.}
\newblock \bibinfo{title}{Universal critical dynamics in high resolution
  neuronal avalanche data}.
\newblock \emph{\bibinfo{journal}{Phys Rev Lett}}
  \textbf{\bibinfo{volume}{108}}, \bibinfo{pages}{208102}
  (\bibinfo{year}{2012}).

\bibitem{Gireesh08}
\bibinfo{author}{Gireesh, E.~D.} \& \bibinfo{author}{Plenz, D.}
\newblock \bibinfo{title}{Neuronal avalanches organize as nested theta-and
  beta/gamma-oscillations during development of cortical layer 2/3}.
\newblock \emph{\bibinfo{journal}{P Natl Acad Sci USA}}
  \textbf{\bibinfo{volume}{105}}, \bibinfo{pages}{7576--7581}
  (\bibinfo{year}{2008}).

\bibitem{Petermann09}
\bibinfo{author}{Petermann, T.} \emph{et~al.}
\newblock \bibinfo{title}{{Spontaneous cortical activity in awake monkeys
  composed of neuronal avalanches.}}
\newblock \emph{\bibinfo{journal}{P Natl Acad Sci USA}}
  \textbf{\bibinfo{volume}{106}}, \bibinfo{pages}{15921--15926}
  (\bibinfo{year}{2009}).

\bibitem{Ribeiro10}
\bibinfo{author}{Ribeiro, T.~L.} \emph{et~al.}
\newblock \bibinfo{title}{Spike avalanches exhibit universal dynamics across
  the sleep-wake cycle}.
\newblock \emph{\bibinfo{journal}{{PL}o{S} ONE}} \textbf{\bibinfo{volume}{5}},
  \bibinfo{pages}{e14129} (\bibinfo{year}{2010}).

\bibitem{Miller09}
\bibinfo{author}{Miller, K.~J.}, \bibinfo{author}{Sorensen, L.~B.},
  \bibinfo{author}{Ojemann, J.~G.} \& \bibinfo{author}{Den~Nijs, M.}
\newblock \bibinfo{title}{Power-law scaling in the brain surface electric
  potential}.
\newblock \emph{\bibinfo{journal}{{PL}o{S} Comput Biol}}
  \textbf{\bibinfo{volume}{5}}, \bibinfo{pages}{e1000609}
  (\bibinfo{year}{2009}).

\bibitem{Linkenkaer01}
\bibinfo{author}{Linkenkaer-Hansen, K.}, \bibinfo{author}{Nikouline, V.~V.},
  \bibinfo{author}{Palva, J.~M.} \& \bibinfo{author}{Ilmoniemi, R.~J.}
\newblock \bibinfo{title}{Long-range temporal correlations and scaling behavior
  in human brain oscillations.}
\newblock \emph{\bibinfo{journal}{J Neurosci}} \textbf{\bibinfo{volume}{21}},
  \bibinfo{pages}{1370--7} (\bibinfo{year}{2001}).

\bibitem{Kitzbichler09}
\bibinfo{author}{Kitzbichler, M.~G.}, \bibinfo{author}{Smith, M.~L.},
  \bibinfo{author}{Christensen, S.~R.} \& \bibinfo{author}{Bullmore, E.}
\newblock \bibinfo{title}{Broadband criticality of human brain network
  synchronization}.
\newblock \emph{\bibinfo{journal}{{PL}o{S} Comput Biol}}
  \textbf{\bibinfo{volume}{5}}, \bibinfo{pages}{e1000314}
  (\bibinfo{year}{2009}).

\bibitem{Eguiluz05}
\bibinfo{author}{Eguiluz, V.~M.}, \bibinfo{author}{Chialvo, D.~R.},
  \bibinfo{author}{Cecchi, G.~A.}, \bibinfo{author}{Baliki, M.} \&
  \bibinfo{author}{Apkarian, A.~V.}
\newblock \bibinfo{title}{Scale-free brain functional networks.}
\newblock \emph{\bibinfo{journal}{Phys Rev Lett}}
  \textbf{\bibinfo{volume}{94}}, \bibinfo{pages}{4} (\bibinfo{year}{2005}).

\bibitem{Sporns10}
\bibinfo{author}{Sporns, O.}
\newblock \emph{\bibinfo{title}{Networks of the Brain}}
  (\bibinfo{publisher}{MIT Press}, \bibinfo{year}{2010}).

\bibitem{Kinouchi06}
\bibinfo{author}{Kinouchi, O.} \& \bibinfo{author}{Copelli, M.}
\newblock \bibinfo{title}{Optimal dynamical range of excitable networks at
  criticality}.
\newblock \emph{\bibinfo{journal}{Nat Phys}} \textbf{\bibinfo{volume}{2}},
  \bibinfo{pages}{348--351} (\bibinfo{year}{2006}).

\bibitem{Shew09}
\bibinfo{author}{Shew, W.}, \bibinfo{author}{Yang, H.},
  \bibinfo{author}{Petermann, T.}, \bibinfo{author}{Roy, R.} \&
  \bibinfo{author}{Plenz, D.}
\newblock \bibinfo{title}{Neuronal avalanches imply maximum dynamic range in
  cortical networks at criticality}.
\newblock \emph{\bibinfo{journal}{J Neurosci}} \textbf{\bibinfo{volume}{29}},
  \bibinfo{pages}{15595--15600} (\bibinfo{year}{2009}).

\bibitem{Haldeman05}
\bibinfo{author}{Haldeman, C.} \& \bibinfo{author}{Beggs, J.~M.}
\newblock \bibinfo{title}{Critical branching captures activity in living neural
  networks and maximizes the number of metastable states}.
\newblock \emph{\bibinfo{journal}{Phys Rev Lett}}
  \textbf{\bibinfo{volume}{94}}, \bibinfo{pages}{058101}
  (\bibinfo{year}{2005}).

\bibitem{Shew11}
\bibinfo{author}{Shew, W.~L.}, \bibinfo{author}{Yang, H.}, \bibinfo{author}{Yu,
  S.}, \bibinfo{author}{Roy, R.} \& \bibinfo{author}{Plenz, D.}
\newblock \bibinfo{title}{Information capacity and transmission are maximized
  in balanced cortical networks with neuronal avalanches}.
\newblock \emph{\bibinfo{journal}{J Neurosci}} \textbf{\bibinfo{volume}{31}},
  \bibinfo{pages}{55--63} (\bibinfo{year}{2011}).

\bibitem{Wen09}
\bibinfo{author}{Wen, Q.}, \bibinfo{author}{Stepanyants, A.},
  \bibinfo{author}{Elston, G.~N.}, \bibinfo{author}{Grosberg, A.~Y.} \&
  \bibinfo{author}{Chklovskii, D.~B.}
\newblock \bibinfo{title}{Maximization of the connectivity repertoire as a
  statistical principle governing the shapes of dendritic arbors.}
\newblock \emph{\bibinfo{journal}{P Natl Acad Sci USA}}
  \textbf{\bibinfo{volume}{106}}, \bibinfo{pages}{12536--12541}
  (\bibinfo{year}{2009}).

\bibitem{Zomorrodi10}
\bibinfo{author}{Zomorrodi, R.}, \bibinfo{author}{Ferecsk\'{o}, A.~S.},
  \bibinfo{author}{Kov\'{a}cs, K.}, \bibinfo{author}{Kr\"{o}ger, H.} \&
  \bibinfo{author}{Timofeev, I.}
\newblock \bibinfo{title}{Analysis of morphological features of thalamocortical
  neurons from the ventroposterolateral nucleus of the cat.}
\newblock \emph{\bibinfo{journal}{J Comp Neurol}}
  \textbf{\bibinfo{volume}{518}}, \bibinfo{pages}{3541--3556}
  (\bibinfo{year}{2010}).

\bibitem{Snider10}
\bibinfo{author}{Snider, J.}, \bibinfo{author}{Pillai, A.} \&
  \bibinfo{author}{Stevens, C.~F.}
\newblock \bibinfo{title}{A universal property of axonal and dendritic arbors.}
\newblock \emph{\bibinfo{journal}{Neuron}} \textbf{\bibinfo{volume}{66}},
  \bibinfo{pages}{45--56} (\bibinfo{year}{2010}).

\bibitem{Koch}
\bibinfo{author}{Koch, C.}
\newblock \emph{\bibinfo{title}{Biophysics of Computation}}
  (\bibinfo{publisher}{Oxford University Press}, \bibinfo{address}{New York},
  \bibinfo{year}{1999}).

\bibitem{Reyes01}
\bibinfo{author}{Reyes, A.}
\newblock \bibinfo{title}{Influence of dendritic conductances on the
  input-output properties of neurons.}
\newblock \emph{\bibinfo{journal}{Annu Rev Neurosci}}
  \textbf{\bibinfo{volume}{24}}, \bibinfo{pages}{653--675}
  (\bibinfo{year}{2001}).

\bibitem{Johnston08}
\bibinfo{author}{Johnston, D.} \& \bibinfo{author}{Narayanan, R.}
\newblock \bibinfo{title}{{Active dendrites: colorful wings of the mysterious
  butterflies.}}
\newblock \emph{\bibinfo{journal}{Trends Neurosci}}
  \textbf{\bibinfo{volume}{31}}, \bibinfo{pages}{309--316}
  (\bibinfo{year}{2008}).

\bibitem{Coop10}
\bibinfo{author}{Coop, A.~D.}, \bibinfo{author}{Cornelis, H.} \&
  \bibinfo{author}{Santamaria, F.}
\newblock \bibinfo{title}{Dendritic excitability modulates dendritic
  information processing in a purkinje cell model}.
\newblock \emph{\bibinfo{journal}{Front Comput Neurosci}}
  \textbf{\bibinfo{volume}{4}}, \bibinfo{pages}{10} (\bibinfo{year}{2010}).

\bibitem{Rall64}
\bibinfo{author}{Rall, W.}
\newblock \bibinfo{title}{Theoretical significance of dendritic trees for
  neuronal input-output relations}.
\newblock In \bibinfo{editor}{Reiss, R.~F.} (ed.)
  \emph{\bibinfo{booktitle}{Neural Theory and Modeling}}
  (\bibinfo{publisher}{Stanford Univ. Press}, \bibinfo{address}{Stanford, CA},
  \bibinfo{year}{1964}).

\bibitem{Neuron}
\bibinfo{author}{Carnevale, N.~T.} \& \bibinfo{author}{Hines, M.~L.}
\newblock \emph{\bibinfo{title}{The {NEURON} Book}}
  (\bibinfo{publisher}{Cambridge University Press}, \bibinfo{year}{2009}).

\bibitem{Stuart99}
\bibinfo{editor}{Stuart, G.}, \bibinfo{editor}{Spruston, N.} \&
  \bibinfo{editor}{H\"ausser, M.} (eds.) \emph{\bibinfo{title}{Dendrites}}
  (\bibinfo{publisher}{Oxford University Press}, \bibinfo{address}{New York},
  \bibinfo{year}{1999}).

\bibitem{London05}
\bibinfo{author}{London, M.} \& \bibinfo{author}{H\"{a}usser, M.}
\newblock \bibinfo{title}{Dendritic computation}.
\newblock \emph{\bibinfo{journal}{Annu Rev Neurosci}}
  \textbf{\bibinfo{volume}{28}}, \bibinfo{pages}{503--532}
  (\bibinfo{year}{2005}).

\bibitem{Marro99}
\bibinfo{author}{Marro, J.} \& \bibinfo{author}{Dickman, R.}
\newblock \emph{\bibinfo{title}{Nonequilibrium Phase Transition in Lattice
  Models}} (\bibinfo{publisher}{Cambridge University Press},
  \bibinfo{address}{Cambridge}, \bibinfo{year}{1999}).

\bibitem{Antic10}
\bibinfo{author}{Antic, S.~D.}, \bibinfo{author}{Zhou, W.-L.},
  \bibinfo{author}{Moore, A.~R.}, \bibinfo{author}{Short, S.~M.} \&
  \bibinfo{author}{Ikonomu, K.~D.}
\newblock \bibinfo{title}{The decade of the dendritic {NMDA} spike.}
\newblock \emph{\bibinfo{journal}{J Neurosci Res}}
  \textbf{\bibinfo{volume}{3001}}, \bibinfo{pages}{2991--3001}
  (\bibinfo{year}{2010}).

\bibitem{Stevens75}
\bibinfo{author}{Stevens, S.~S.}
\newblock \emph{\bibinfo{title}{Psychophysics: Introduction to its perceptual,
  neural, and social prospects}} (\bibinfo{publisher}{John Wiley and Sons},
  \bibinfo{year}{1975}).

\bibitem{BakChialvo03}
\bibinfo{author}{Bak, P.} \& \bibinfo{author}{Chialvo, D.~R.}
\newblock \bibinfo{title}{Adaptive learning by extremal dynamics and negative
  feedback}.
\newblock \emph{\bibinfo{journal}{Phys Rev E}} \textbf{\bibinfo{volume}{63}},
  \bibinfo{pages}{031912} (\bibinfo{year}{2001}).

\bibitem{Levina07}
\bibinfo{author}{Levina, A.}, \bibinfo{author}{Herrmann, J.~M.} \&
  \bibinfo{author}{Geisel, T.}
\newblock \bibinfo{title}{Dynamical synapses causing self-organized criticality
  in neural networks}.
\newblock \emph{\bibinfo{journal}{Nat Phys}} \textbf{\bibinfo{volume}{3}},
  \bibinfo{pages}{857--860} (\bibinfo{year}{2007}).

\bibitem{Rubinov11}
\bibinfo{author}{Rubinov, M.}, \bibinfo{author}{Sporns, O.},
  \bibinfo{author}{Thivierge, J.-P.} \& \bibinfo{author}{Breakspear, M.}
\newblock \bibinfo{title}{Neurobiologically realistic determinants of
  self-organized criticality in networks of spiking neurons}.
\newblock \emph{\bibinfo{journal}{{PL}o{S} Comput Biol}}
  \textbf{\bibinfo{volume}{7}}, \bibinfo{pages}{e1002038}
  (\bibinfo{year}{2011}).

\bibitem{Gal12}
\bibinfo{author}{Gal, A.} \& \bibinfo{author}{Marom, S.}
\newblock \bibinfo{title}{Self-organized criticality in single neuron
  excitability}.
\newblock \emph{\bibinfo{journal}{arXiv preprint arXiv:1210.7414}}
  (\bibinfo{year}{2012}).

\bibitem{Gollo12b}
\bibinfo{author}{Gollo, L.~L.}, \bibinfo{author}{Mirasso, C.} \&
  \bibinfo{author}{Egu{\'\i}luz, V.~M.}
\newblock \bibinfo{title}{Signal integration enhances the dynamic range in
  neuronal systems}.
\newblock \emph{\bibinfo{journal}{Phys Rev E}} \textbf{\bibinfo{volume}{85}},
  \bibinfo{pages}{040902} (\bibinfo{year}{2012}).

\bibitem{Gollo09}
\bibinfo{author}{Gollo, L.~L.}, \bibinfo{author}{Kinouchi, O.} \&
  \bibinfo{author}{Copelli, M.}
\newblock \bibinfo{title}{Active dendrites enhance neuronal dynamic range}.
\newblock \emph{\bibinfo{journal}{{PL}o{S} Comput Biol}}
  \textbf{\bibinfo{volume}{5}}, \bibinfo{pages}{e1000402}
  (\bibinfo{year}{2009}).

\bibitem{Gollo12}
\bibinfo{author}{Gollo, L.~L.}, \bibinfo{author}{Kinouchi, O.} \&
  \bibinfo{author}{Copelli, M.}
\newblock \bibinfo{title}{Statistical physics approach to dendritic
  computation: The excitable-wave mean-field approximation}.
\newblock \emph{\bibinfo{journal}{Phys Rev E}} \textbf{\bibinfo{volume}{85}},
  \bibinfo{pages}{011911} (\bibinfo{year}{2012}).

\bibitem{Assis08}
\bibinfo{author}{Assis, V. R.~V.} \& \bibinfo{author}{Copelli, M.}
\newblock \bibinfo{title}{Dynamic range of hypercubic stochastic excitable
  media}.
\newblock \emph{\bibinfo{journal}{Phys Rev E}} \textbf{\bibinfo{volume}{77}},
  \bibinfo{pages}{011923} (\bibinfo{year}{2008}).

\bibitem{Larremore11}
\bibinfo{author}{Larremore, D.}, \bibinfo{author}{Shew, W.} \&
  \bibinfo{author}{Restrepo, J.}
\newblock \bibinfo{title}{{Predicting Criticality and Dynamic Range in Complex
  Networks: Effects of Topology}}.
\newblock \emph{\bibinfo{journal}{Phys Rev Lett}}
  \textbf{\bibinfo{volume}{106}}, \bibinfo{pages}{058101}
  (\bibinfo{year}{2011}).

\bibitem{Llinas80}
\bibinfo{author}{Llin\'{a}s, R.} \& \bibinfo{author}{Sugimori, M.}
\newblock \bibinfo{title}{Electrophysiological properties of in vitro purkinje
  cell dendrites in mammalian cerebellar slices.}
\newblock \emph{\bibinfo{journal}{J Physiol}} \textbf{\bibinfo{volume}{305}},
  \bibinfo{pages}{197--213} (\bibinfo{year}{1980}).

\bibitem{Schiller97}
\bibinfo{author}{Schiller, J.}, \bibinfo{author}{Schiller, Y.},
  \bibinfo{author}{Stuart, G.} \& \bibinfo{author}{Sakmann, B.}
\newblock \bibinfo{title}{Calcium action potentials restricted to distal apical
  dendrites of rat neocortical pyramidal neurons.}
\newblock \emph{\bibinfo{journal}{J Physiol}} \textbf{\bibinfo{volume}{505}},
  \bibinfo{pages}{605--616} (\bibinfo{year}{1997}).

\bibitem{Manchanda13}
\bibinfo{author}{Manchanda, K.}, \bibinfo{author}{Yadav, A.~C.} \&
  \bibinfo{author}{Ramaswamy, R.}
\newblock \bibinfo{title}{Scaling behavior in probabilistic neuronal cellular
  automata}.
\newblock \emph{\bibinfo{journal}{Phys Rev E}} \textbf{\bibinfo{volume}{87}},
  \bibinfo{pages}{012704} (\bibinfo{year}{2013}).

\bibitem{Villacorta13}
\bibinfo{author}{Villacorta-Atienza, J.~A.} \& \bibinfo{author}{Makarov, V.~A.}
\newblock \bibinfo{title}{Wave-processing of long-scale information by neuronal
  chains}.
\newblock \emph{\bibinfo{journal}{{PL}o{S} ONE}} \textbf{\bibinfo{volume}{8}},
  \bibinfo{pages}{e57440} (\bibinfo{year}{2013}).

\bibitem{Suzuki08}
\bibinfo{author}{Suzuki, T.}, \bibinfo{author}{Kodama, S.},
  \bibinfo{author}{Hoshino, C.}, \bibinfo{author}{Izumi, T.} \&
  \bibinfo{author}{Miyakawa, H.}
\newblock \bibinfo{title}{A plateau potential mediated by the activation of
  extrasynaptic {NMDA} receptors in rat hippocampal {CA}1 pyramidal neurons.}
\newblock \emph{\bibinfo{journal}{Eur J Neurosci}}
  \textbf{\bibinfo{volume}{28}}, \bibinfo{pages}{521--534}
  (\bibinfo{year}{2008}).

\bibitem{Davie08}
\bibinfo{author}{Davie, J.~T.}, \bibinfo{author}{Clark, B.~A.} \&
  \bibinfo{author}{H\"{a}usser, M.}
\newblock \bibinfo{title}{The origin of the complex spike in cerebellar
  purkinje cells.}
\newblock \emph{\bibinfo{journal}{J Neurosci}} \textbf{\bibinfo{volume}{28}},
  \bibinfo{pages}{7599--7609} (\bibinfo{year}{2008}).

\bibitem{Major08}
\bibinfo{author}{Major, G.}, \bibinfo{author}{Polsky, A.},
  \bibinfo{author}{Denk, W.}, \bibinfo{author}{Schiller, J.} \&
  \bibinfo{author}{Tank, D.~W.}
\newblock \bibinfo{title}{Spatiotemporally graded {NMDA} spike/plateau
  potentials in basal dendrites of neocortical pyramidal neurons.}
\newblock \emph{\bibinfo{journal}{J Neurophysiol}}
  \textbf{\bibinfo{volume}{99}}, \bibinfo{pages}{2584--2601}
  (\bibinfo{year}{2008}).

\bibitem{Wong79}
\bibinfo{author}{Wong, R.~K.} \& \bibinfo{author}{Prince, D.~A.}
\newblock \bibinfo{title}{Dendritic mechanisms underlying penicillin-induced
  epileptiform activity.}
\newblock \emph{\bibinfo{journal}{Science}} \textbf{\bibinfo{volume}{204}},
  \bibinfo{pages}{1228--1231} (\bibinfo{year}{1979}).

\bibitem{Wong92}
\bibinfo{author}{Wong, R.~K.} \& \bibinfo{author}{Stewart, M.}
\newblock \bibinfo{title}{Different firing patterns generated in dendrites and
  somata of {CA}1 pyramidal neurones in guinea-pig hippocampus.}
\newblock \emph{\bibinfo{journal}{J Physiol}} \textbf{\bibinfo{volume}{457}},
  \bibinfo{pages}{675--687} (\bibinfo{year}{1992}).

\bibitem{Kamondi98}
\bibinfo{author}{Kamondi, A.}, \bibinfo{author}{Acs\'{a}dy, L.},
  \bibinfo{author}{Wang, X.~J.} \& \bibinfo{author}{Buzs\'{a}ki, G.}
\newblock \bibinfo{title}{Theta oscillations in somata and dendrites of
  hippocampal pyramidal cells in vivo: activity-dependent phase-precession of
  action potentials.}
\newblock \emph{\bibinfo{journal}{Hippocampus}} \textbf{\bibinfo{volume}{8}},
  \bibinfo{pages}{244–61} (\bibinfo{year}{1998}).

\bibitem{Remme09}
\bibinfo{author}{Remme, M. W.~H.}, \bibinfo{author}{Lengyel, M.} \&
  \bibinfo{author}{Gutkin, B.~S.}
\newblock \bibinfo{title}{The role of ongoing dendritic oscillations in
  single-neuron dynamics}.
\newblock \emph{\bibinfo{journal}{{PL}o{S} Comput Biol}}
  \textbf{\bibinfo{volume}{5}}, \bibinfo{pages}{e1000493}
  (\bibinfo{year}{2009}).

\bibitem{Branco10}
\bibinfo{author}{Branco, T.} \& \bibinfo{author}{H{\"a}usser, M.}
\newblock \bibinfo{title}{The single dendritic branch as a fundamental
  functional unit in the nervous system}.
\newblock \emph{\bibinfo{journal}{Curr Opin Neurobiol}}
  \textbf{\bibinfo{volume}{20}}, \bibinfo{pages}{494--502}
  (\bibinfo{year}{2010}).

\bibitem{Carelli05}
\bibinfo{author}{Carelli, P.~V.}, \bibinfo{author}{Reyes, M.~B.},
  \bibinfo{author}{Sartorelli, J.~C.} \& \bibinfo{author}{Pinto, R.~D.}
\newblock \bibinfo{title}{Whole cell stochastic model reproduces the
  irregularities found in the membrane potential of bursting neurons}.
\newblock \emph{\bibinfo{journal}{J Neurophysiol}}
  \textbf{\bibinfo{volume}{94}}, \bibinfo{pages}{1169--1179}
  (\bibinfo{year}{2005}).

\bibitem{Cannon10}
\bibinfo{author}{Cannon, R.~C.}, \bibinfo{author}{O'Donnell, C.} \&
  \bibinfo{author}{Nolan, M.~F.}
\newblock \bibinfo{title}{Stochastic ion channel gating in dendritic neurons:
  morphology dependence and probabilistic synaptic activation of dendritic
  spikes}.
\newblock \emph{\bibinfo{journal}{{PL}o{S} Comput Biol}}
  \textbf{\bibinfo{volume}{6}}, \bibinfo{pages}{e1000886}
  (\bibinfo{year}{2010}).

\bibitem{Jarsky05}
\bibinfo{author}{Jarsky, T.}, \bibinfo{author}{Roxin, A.},
  \bibinfo{author}{Kath, W.~L.} \& \bibinfo{author}{Spruston, N.}
\newblock \bibinfo{title}{Conditional dendritic spike propagation following
  distal synaptic activation of hippocampal {CA}1 pyramidal neurons.}
\newblock \emph{\bibinfo{journal}{Nat Neurosci}} \textbf{\bibinfo{volume}{8}},
  \bibinfo{pages}{1667--1676} (\bibinfo{year}{2005}).

\bibitem{Publio12}
\bibinfo{author}{Publio, R.}, \bibinfo{author}{Ceballos, C.~C.} \&
  \bibinfo{author}{Roque, A.~C.}
\newblock \bibinfo{title}{Dynamic range of vertebrate retina ganglion cells:
  Importance of active dendrites and coupling by electrical synapses}.
\newblock \emph{\bibinfo{journal}{PloS ONE}} \textbf{\bibinfo{volume}{7}},
  \bibinfo{pages}{e48517} (\bibinfo{year}{2012}).

\bibitem{Copelli02}
\bibinfo{author}{Copelli, M.}, \bibinfo{author}{Roque, A.~C.},
  \bibinfo{author}{Oliveira, R.~F.} \& \bibinfo{author}{Kinouchi, O.}
\newblock \bibinfo{title}{Physics of {P}sychophysics: {S}tevens and
  {W}eber-{F}echner laws are transfer functions of excitable media}.
\newblock \emph{\bibinfo{journal}{Phys Rev E}} \textbf{\bibinfo{volume}{65}},
  \bibinfo{pages}{060901} (\bibinfo{year}{2002}).

\bibitem{Furtado06}
\bibinfo{author}{Furtado, L.~S.} \& \bibinfo{author}{Copelli, M.}
\newblock \bibinfo{title}{Response of electrically coupled spiking neurons: a
  cellular automaton approach}.
\newblock \emph{\bibinfo{journal}{Phys Rev E}} \textbf{\bibinfo{volume}{73}},
  \bibinfo{pages}{011907} (\bibinfo{year}{2006}).

\bibitem{Ribeiro08}
\bibinfo{author}{Ribeiro, T.~L.} \& \bibinfo{author}{Copelli, M.}
\newblock \bibinfo{title}{Deterministic excitable media under {P}oisson drive:
  Power law responses, spiral waves and dynamic range}.
\newblock \emph{\bibinfo{journal}{Phys Rev E}} \textbf{\bibinfo{volume}{77}},
  \bibinfo{pages}{051911} (\bibinfo{year}{2008}).

\bibitem{Publio09}
\bibinfo{author}{Publio, R.}, \bibinfo{author}{Oliveira, R.~F.} \&
  \bibinfo{author}{Roque, A.~C.}
\newblock \bibinfo{title}{A computational study on the role of gap junctions
  and rod {Ih} conductance in the enhancement of the dynamic range of the
  retina}.
\newblock \emph{\bibinfo{journal}{{PL}o{S} ONE}} \textbf{\bibinfo{volume}{4}},
  \bibinfo{pages}{e6970} (\bibinfo{year}{2009}).

\bibitem{Jan10}
\bibinfo{author}{Jan, Y.-N.} \& \bibinfo{author}{Jan, L.~Y.}
\newblock \bibinfo{title}{Branching out: mechanisms of dendritic arborization}.
\newblock \emph{\bibinfo{journal}{Nat Rev Neurosci}}
  \textbf{\bibinfo{volume}{11}}, \bibinfo{pages}{316--328}
  (\bibinfo{year}{2010}).

\bibitem{Woolley90}
\bibinfo{author}{Woolley, C.~S.}, \bibinfo{author}{Gould, E.},
  \bibinfo{author}{Frankfurt, M.} \& \bibinfo{author}{McEwen, B.~S.}
\newblock \bibinfo{title}{Naturally occurring fluctuation in dendritic spine
  density on adult hippocampal pyramidal neurons}.
\newblock \emph{\bibinfo{journal}{J Neurosci}} \textbf{\bibinfo{volume}{10}},
  \bibinfo{pages}{4035--4039} (\bibinfo{year}{1990}).

\bibitem{Bonachela10}
\bibinfo{author}{Bonachela, J.~A.}, \bibinfo{author}{{De Franciscis}, S.},
  \bibinfo{author}{Torres, J.~J.} \& \bibinfo{author}{Mu\~{n}oz, M.~A.}
\newblock \bibinfo{title}{{Self-organization without conservation: {A}re
  neuronal avalanches generically critical?}}
\newblock \emph{\bibinfo{journal}{J Stat Mech-Theory E}}
  \textbf{\bibinfo{volume}{2010}}, \bibinfo{pages}{28} (\bibinfo{year}{2010}).

\bibitem{Parish04}
\bibinfo{author}{Parish, L.~M.} \emph{et~al.}
\newblock \bibinfo{title}{Long-range temporal correlations in epileptogenic and
  non-epileptogenic human hippocampus.}
\newblock \emph{\bibinfo{journal}{Neuroscience}}
  \textbf{\bibinfo{volume}{125}}, \bibinfo{pages}{1069--76}
  (\bibinfo{year}{2004}).

\bibitem{Bhattacharya05}
\bibinfo{author}{Bhattacharya, J.}, \bibinfo{author}{Edwards, J.},
  \bibinfo{author}{Mamelak, A.~N.} \& \bibinfo{author}{Schuman, E.~M.}
\newblock \bibinfo{title}{Long-range temporal correlations in the spontaneous
  spiking of neurons in the hippocampal-amygdala complex of humans.}
\newblock \emph{\bibinfo{journal}{Neuroscience}}
  \textbf{\bibinfo{volume}{131}}, \bibinfo{pages}{547--555}
  (\bibinfo{year}{2005}).

\bibitem{Mazzoni07}
\bibinfo{author}{Mazzoni, A.} \emph{et~al.}
\newblock \bibinfo{title}{On the dynamics of the spontaneous activity in
  neuronal networks}.
\newblock \emph{\bibinfo{journal}{{PL}o{S} ONE}} \textbf{\bibinfo{volume}{2}},
  \bibinfo{pages}{e439} (\bibinfo{year}{2007}).

\end{thebibliography}

%%
%% TABLES
%%
%% If there are any tables, put them here.
%%

\end{document}